\documentclass[a4paper,12pt]{article}
\usepackage{amsmath}
\usepackage{amsfonts}
\usepackage{authblk}
\usepackage{graphicx,epsfig}
\usepackage{float}
\usepackage{subcaption}  
\usepackage{amssymb}
\usepackage{color}

\numberwithin{equation}{section}
\title{Bound on Rindler trajectories in Black Hole spacetime}

\author{Kajol Paithankar\footnote{kajol.paithankar@cbs.ac.in}}
\author{Sanved Kolekar\footnote{sanved.kolekar@cbs.ac.in} \,}
\affil{ UM-DAE Centre for Excellence in Basic Sciences,\\  
Mumbai 400098, India}

\date{January 2019}
\begin{document}

\maketitle

\begin{abstract}

We investigate radial Rindler trajectories in a static spherically symmetric black hole spacetime. We assume the trajectory to remain linearly uniformly accelerated throughout its motion, in the sense of the curved spacetime generalisation of the Letaw-Frenet equations. For the Schwarzschild spacetime, we arrive at a bound on the magnitude of the acceleration $|a|$ for radially inward moving trajectories, in terms of the mass $M$ of the black hole given by $|a| \leq 1/(\sqrt{27} M)$ for a particular choice of asymptotic initial data $h$, such that, for acceleration $|a|$ greater than the bound value, the linearly uniformly accelerated  trajectory always falls into the black hole. For $|a|$ satisfying the bound, there is a minimum radius or the distance of closest approach for the radial linearly uniformly accelerated trajectory to escape back to infinity. However, this distance of closest approach is found to approach its lowest value of $r_b = 3M $, greater than the Schwarzschild radius of the black hole, when the bound, $|a| = 1/( \sqrt{27}M)$ is saturated. We further show that a finite bound on the value of acceleration, $ |a| \leq \mathcal{B}(M,h)$ and a corresponding distance of closest approach $r_{b} > 2M$ always exists, for all finite asymptotic initial data $h$.

\end{abstract}

\section{Introduction:}\label{sec1}

Rindler trajectories are a special class of trajectories in flat spacetime. Hyperbolic motion along the orbits of the boost Killing vector lie in the right quadrant of the Minkowski spacetime constrained by the $X = -T$ and $X = T$ null surfaces where $X$ and $T$ are the Minkowski co-ordinates. This special structure together with the constancy in the magnitude of acceleration $|a|$ leads to a notion of temperature $T_u$ equal to $|a|/2 \pi$ associated with the Killing horizon $X = T$ known as the Unruh effect when one analyses the different vacua associated with the background quantum fields in the global inertial frame and the Rindler frame \cite{davies, unruh}. It has been argued that in a general curved spacetime one can construct, in principle, trajectories which are locally Rindler and associate a first law of thermodynamics with the corresponding local Rindler horizon by analysing the flow of matter flux through it. \cite{local1, local2, local3, local4, local5}.

\begin{figure}[h]
\centering
\includegraphics[width=10cm,height=8cm]{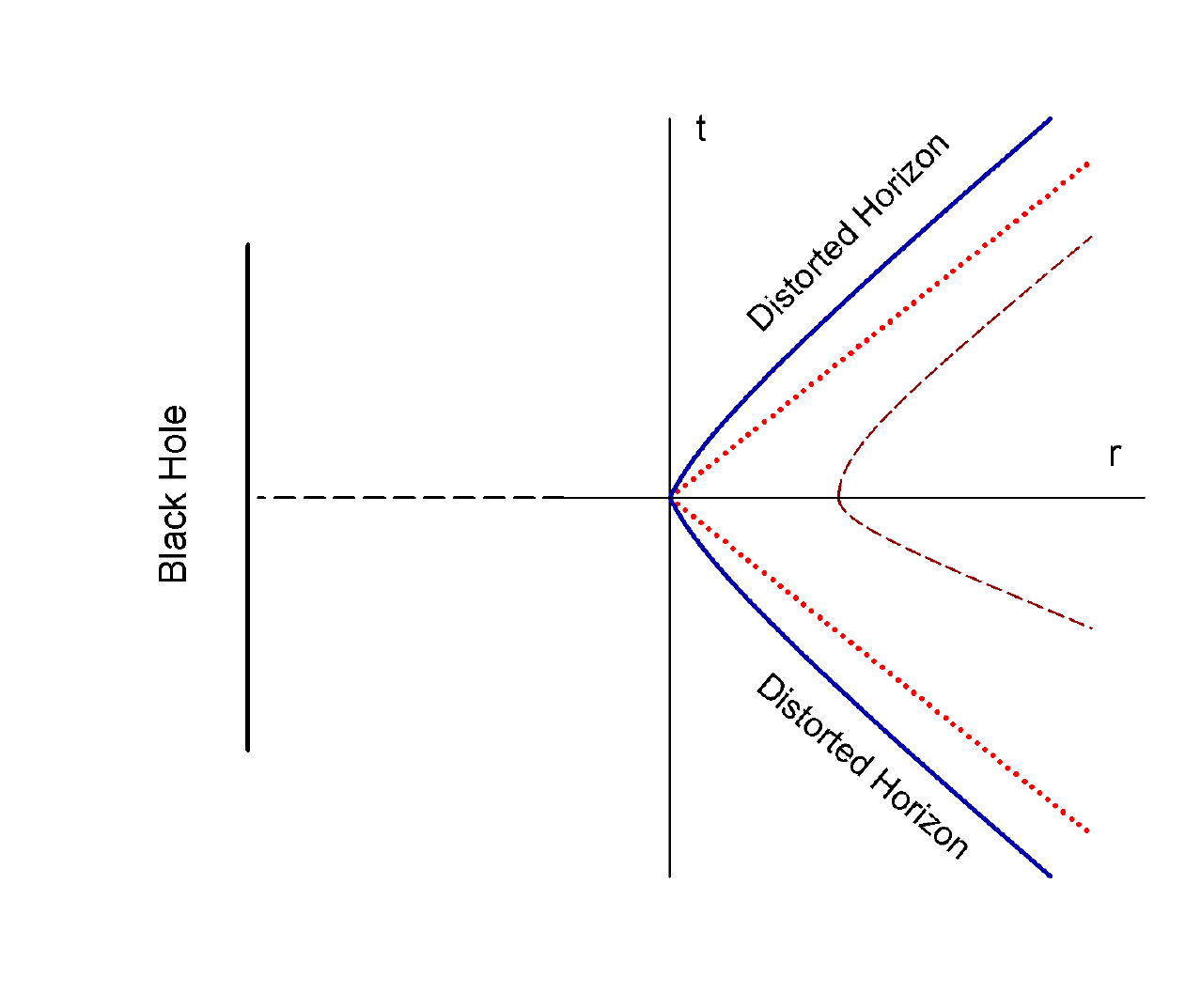}
\caption{Minkowski-Rindler setting perturbed by Black Hole}
\label{perturbedhorizon}
\end{figure}

Suppose one introduces a static spherically symmetric black hole, of the Schwarzschild type with mass $M$, in the Minkowski-Rindler setting such that the black hole is at a large distance away from the Rindler observer. The spacetime geometry in the local neighbourhood of the Rindler trajectory will still be dominantly flat but now also have small perturbations to the background flat metric due to the presence of the black hole. The hyperbolic solution to the uniformly accelerated trajectory and the corresponding Rindler horizon null surface too would obtain corrections, say, to linear order in the metric perturbations. However one can expect the broader quadrant structure formed by the perturbed $X = T$ null surface (casual past of the future asymptotic point), the perturbed $X = -T$ null surface (casual future of the past asymptotic point), the past null infinity and future null infinity of the Rindler spacetime to deform from the usual rectangular shape since the null trajectories are no longer Minkowski straight lines in the $(X,T)$ plane due to the metric perturbation (see Figure \ref{perturbedhorizon}). In the present picture, due to the presence of the black hole, the $X$ and $T$ co-ordinates behave like the $r$ and $t$ of the Schwarzschild co-ordinate while the the $Y$ and $Z$ transverse co-ordinates behave like the corresponding angular co-ordinates $\theta$, $\phi$ at $r \rightarrow \infty$. 
Note that the scenario described above is qualitatively different than the familiar $r =$ constant uniformly accelerated observers in the Schwarzschild metric. In the latter, the magnitude of the acceleration is dependant on both the stationary co-ordinate $r$ and mass $M$ of the black hole whereas in the former the acceleration magnitude $|a|$ is an independent parameter and hence the trajectory is not restricted to just a constant $r$. In the present case there are two horizons present, the black hole horizon and the Rindler horizon, whose relative positions are determined by the parameters $M$ and $|a|$ (or a combination of $|a|$ and $M$) respectively. An analysis of the quantum fields in this background metric should lead to interesting consequences for the temperature $T_u$ determined by the accelerated observer with $T_u$ now depending on both the parameters $|a|$ and $M$. 

The more general case, wherein the black hole is not restricted to be far away from the Rindler observer, is even more intriguing.  
Obtaining the corresponding uniformly accelerated trajectory would be highly non-trivial since one can expect strong curvature effects to significantly modify the hyperbolic trajectory.  Additionally, one would like to understand the features of the deformed Rindler horizon, if at all existent, in the presence of the black hole. We focus on the former aspect in the present paper.

In a curved spacetime, a generalisation of the Rindler trajectory involves, in addition to the constancy condition on the magnitude of acceleration, a further constraint on the trajectory that it must be \textit{linear} with vanishing torsion and hyper-torsion \cite{rindler}. To be consistent, the linear uniformly accelerating trajectory should reduce to the usual hyperbolic trajectory in a local inertial frame defined around any event along the trajectory. In \cite{kolekar}, a construction based on the Letaw-Frenet equations and their corresponding geometrical scalar invariants was shown to lead to such a covariant definition of the linear uniformly accelerated (LUA) trajectory. One defines a tetrad of basis vectors $V^i_{\alpha}$ along each point on the trajectory with two of these vectors to lie along the tangent vector $u^i$ and the unit acceleration vector $a^i/|a|$ respectively while the remaining two are defined using the Gram-Schmidt orthogonalisation procedure. The Letaw-Frenet equations in curved spacetime are then $\; u^{j}\nabla_{j}{V_{\alpha}^{i}}=K^{\beta}_{\alpha}V^{i}_{\beta}\;$,  where,
\[K_{\alpha\beta}=
 \left( {\begin{array}{cccc}
  0 & -\mathcal{K(\tau)} & 0 & 0 \\
  \mathcal{K(\tau)} & 0 & -\mathcal{T(\tau)} & 0 \\
  0 & \mathcal{T(\tau)} & 0 & -\mathcal{V(\tau)} \\
  0 & 0 & \mathcal{V(\tau)} & 0 \\ 
 \end{array} } \right)
\]
with $\mathcal{K(\tau)}$, $\mathcal{T(\tau)}$ and $\mathcal{V(\tau)}$ being the scalar invariants of curvature, torsion and hyper-torsion.

The LUA trajectory is then defined as the trajectory with a constant positive curvature and vanishing torsion and hyper-torsion, where the curvature is the magnitude of proper acceleration, $\mathcal{K(\tau)}=|a|$. In the present case, for the LUA trajectory the Letaw-Frenet equation reduce to the following constraint equation
\begin{eqnarray}
w^{i}-{|a|}^{2}u^{i}=0
\label{linearity}
\end{eqnarray}
where $w^{i}=u^{j}{\nabla}_{j}a^{i}$. The solution $x^{i}(\tau)$ consistent with the above constraint equation in a given background curved spacetime is the required trajectory of the linear uniformly accelerated observer, that is the generalised Rindler trajectory.

The paper is organised as follows. In section 2, we set-up the relevant equations and present the general solution for a radial LUA trajectory, in a spherically symmetric background metric of the Schwarzschild type with $g_{tt}=f(r) = -g^{rr}$, in terms of its 4-velocity. Further, the explicit solution for the radial LUA trajectory is obtained for the black hole with $f(r) = 1 - s^2/r^2$. We next apply the results of section 2 to the Schwarzschild case, in section 3, for radial trajectories having acceleration directed away from the black hole. Interestingly, we arrive at a bound on the magnitude of acceleration $|a|$ in terms of the mass $M$ of the black hole given by $ |a| \leq 1/(\sqrt{27} M)$ for a particular choice of asymptotic initial data $h$, such that, for acceleration $|a|$ greater than the bound value, the LUA trajectory always falls into the black hole. For acceleration $|a|$ satisfying the bound, there is a minimum radius or the distance of closest approach for the radial LUA trajectory,
to escape back to infinity. However, this distance of closest approach is found to approach the radius $r_{b} = 3M$, which is greater than the Schwarzschild radius of the black hole, when the value of acceleration saturates the bound, $ |a| = 1/(\sqrt{27} M)$. We further show that a finite bound on the value of acceleration, $ |a| \leq B(M,h)$ and a corresponding distance of closest approach $r_{b} > 2M$ always exists, for all finite asymptotic initial data $h$. In section 4, we find the explicit solution to the LUA trajectory in the de-sitter spacetime and show that the same trajectory viewed in the $5-D$ Minkowski spacetime, embedding the $4-D$ de-sitter spacetime, is also LUA in the $5-D$ sense and vice-versa. The conclusions are presented in section 5.

Latin indices run from 0 to 3 and signature of the metric is taken to be $(+,-,-,-)$.

\section{General Setup:}\label{sec2}

Consider a general metric of the following form for a spherically symmetric and non-rotating black hole, with an asymptotic flat boundary,
\begin{equation}
{ds}^{2} = f(r){dt}^{2}-{f(r)}^{-1}{dr}^{2}-r^{2}{d\theta}^{2}-r^{2}{\sin}^{2}\theta {d\phi}^{2}
\label{2.metric}
\end{equation}
where, $f(r)$ is a smooth differentiable function, with $f(r_{H})=0$ at some radius $r_{H}$ and $f(r)\to1$ as $r\to\infty$.
To find the LUA trajectory in the above background metric, we assume the trajectory to be purely along the radial direction having fixed angular coordinates. That is, for our beginning ansatz, we have $t(\tau)$, $r(\tau)$, $\theta$ =constant and $\phi=$ constant, where $\tau$ is the proper time along the trajectory. The corresponding four velocity $u^{i}$ is then simply,
\begin{equation}
u^{i}=\dfrac{dx^{i}}{d\tau}\equiv\left(u^{0},u^{1},0,0\right)
\label{2.4-velocity}
\end{equation}

The components of acceleration vector $a^{i}=u^{j}{\nabla}_{j}u^{i}$ are
\begin{eqnarray}
a^{0}&=&\dfrac{du^{0}}{d\tau}+\frac{1}{f}\left(\frac{\partial f}{\partial r}\right)u^{0}u^{1}\label{2.acc0}\\
a^{1}&=&\dfrac{du^{1}}{d\tau}+\frac{1}{2}\left(\frac{\partial f}{\partial r}\right)\label{2.acc1}\\
a^{2}&=&{\Gamma^{2}}_{ij}u^{i}u^{j}=0\label{2.acc2}\\
a^{3}&=&{\Gamma^{3}}_{ij}u^{i}u^{j}=0\label{2.acc3}
\end{eqnarray}
For the spherically symmetric form of the metric in Eq.(\ref{2.metric}), the components ${\Gamma^{2}}_{ij}$ and ${\Gamma^{3}}_{ij}$ vanish for $i,j=0,1$. Because of this, the angular components of acceleration vector for a radial trajectory in any metric of the form of Eq.(\ref{2.metric}) are always zero, that is, $a^{2}=a^{3}=0$. The magnitude of proper acceleration is
\begin{eqnarray}
-{|a|}^{2}&=&f(r){(a^{0})}^{2}-\frac{{(a^{1})}^{2}}{f(r)}\nonumber\\
& =& f(r){\left(\dfrac{du^{0}}{d\tau}\right)}^{2}-\frac{1}{f(r)}{\left(\dfrac{du^{1}}{d\tau}\right)}^{2}+\frac{1}{f(r)}{\left(\frac{\partial f}{\partial r}\right)}^{2}{\left(u^{0}u^{1}\right)}^{2}-\frac{1}{4f(r)}{\left(\frac{\partial f}{\partial r}\right)}^{2}\nonumber\\
& &-\frac{1}{f}\left(\frac{\partial f}{\partial r}\right)\left(\dfrac{du^{1}}{d\tau}\right)+2u^{0}u^{1}\left(\frac{\partial f}{\partial r}\right)\left(\dfrac{du^{0}}{d\tau}\right)
\label{2.acc magnitude}
\end{eqnarray}
where we take $|a|$ to be a constant as per our requirement. Further imposing the linearity condition for the LUA trajectory in terms of Eq.(\ref{linearity}), $w^{i}={|a|}^{2}u^{i}$, we have for the time component
\begin{eqnarray}
0 &=&w^{0}-{|a|}^{2}u^{0}=u^{i}{\nabla}_{i}a^{0}-{|a|}^{2}u^{0} \nonumber \\
&=&\dfrac{da^{0}}{d\tau}+\frac{1}{2f} \left( \frac{\partial f}{\partial r} \right) \left(u^{0}a^{1}+a^{0}u^{1}\right)-{|a|}^{2}u^{0}\label{2.linearity0}
\end{eqnarray}
Similarly the equation for radial component becomes
\begin{eqnarray}
0 &=&w^{1}-{|a|}^{2}u^{1}=u^{i}{\nabla}_{i}a^{1}-{|a|}^{2}u^{1} \nonumber \\
 &=&\dfrac{da^{1}}{d\tau}+\frac{1}{2}\dfrac{\partial f}{\partial r}\left(f u^{0}a^{0}-\frac{u^{1}a^{1}}{f}\right)-{|a|}^{2}u^{1}\nonumber \\
 &=&\dfrac{da^{1}}{d\tau}-{|a|}^{2}u^{1}\label{2.linearity1}
\end{eqnarray}
where, the second last term in the above equation vanishes as $u^{i}a_{i}=0$.
The angular components of $w^{i}$, that is, $w^{2}$ and $w^{3}$ again vanish due to spherical symmetry, since the corresponding ${\Gamma^{2}}_{ij}$ and ${\Gamma^{3}}_{ij}$ are zero. 

It is instructive to note here that in accordance to the Letaw- Frenet formulation \cite{kolekar}, the $V_{0}$ and $V_{1}$ tetrad vectors for the spherically symmetric metric in Eq.(\ref{2.metric}) lie in the $(t,r)$ plane for each point along the LUA trajectory, since $u^{i}$ and $a^{i}$ lie completely in the $(t,r)$ plane. Whereas, one can choose the $V_{2}$ and $V_{3}$ tetrad vectors to lie along the $\theta$ and $\phi$ directions respectively. The spherical symmetry further dictates the angular components of $w^{i}$ to vanish. Hence, $w^{i}$ also lies in the $(t,r)$ plane, or in the plane of $(u^{i},a^{i})$ and one can then write $w^{i}= C u^{i}+ D a^{i}$, completely in terms of the basis vectors $V_{0}$ and $V_{1}$. However, since $|a|$ is a constant, one can show that $w^{i}$ is orthogonal to $a^{i}$, that is $w^{i}a_{i}=0$. This can easily be verified starting from,
\[0=u^{j}{\nabla}_{j}(a^{i}a_{i})=2\, w^{i}a_{i}\]
Then for the radial LUA trajectory, in the background metric Eq.(\ref{2.metric}), the condition $w^{i}={|a|}^{2}u^{i}$ is always satisfied. In other words, solving either $|a|^{2}$ = constant or $w^{i}={|a|}^{2}u^{i}$ should be equivalent and lead us to the same solution for $t(\tau)$ and $r(\tau)$.

The radial component of $w^{i}$ is straight forward to solve in the current scenario. We first substitute the expression for $a^{1}$, Eq.(\ref{2.acc1}) in Eq.(\ref{2.linearity1}) to get,
\begin{eqnarray}
\dfrac{d^{2}u^{1}}{d\tau^{2}}&=&\left({|a|}^{2}-\frac{1}{2}\frac{\partial^{2}f}{\partial r^{2}}\right)u^{1}
\end{eqnarray}
Integrating the above equation we get,
\begin{equation}
u^{1}=\pm\sqrt{{|a|}^{2}r^{2}-f(r)+2c_{1}r+2c_{2}}
\label{2.solution1}
\end{equation}
where, $c_{1}$ and $c_{2}$ are constants of integration.
The $u^{0}$ component is obtained trivially through the normalization condition $u^{i}u_{i}=1$, to be,
\begin{equation}
u^{0}={f(r)}^{-1}\sqrt{{|a|}^{2}r^{2}+2c_{1}r+2c_{2}}
\label{2.solution0}
\end{equation}
The above solutions need to be consistent with the $w^{0}={|a|}^{2}u^{0}$ equation in Eq.(\ref{2.linearity0}). Substituting Eq.(\ref{2.solution1}) and Eq.(\ref{2.solution0}) in Eq.(\ref{2.linearity0}) we arrive at the constrain, $c_{2}={c_{1}}^2/2{|a|}^2$.

Collecting our results, we have,
\begin{eqnarray}
u^{1}&=&\pm\sqrt{\left(|a|r+h\right)^2-f(r)}\label{2.result1}\\
u^{0}&=&{f(r)}^{-1}\left(|a|r+h\right)\label{2.result0}
\end{eqnarray}
here $h^2=2c_{2}$ is a constant.
These are the solutions to the Eq.(\ref{2.linearity0}) and Eq.(\ref{2.linearity1}). 
For a given $f(r)$, the trajectory $r(\tau)$ and $t(\tau)$ can be obtained in principle by integrating $u^{1}$ and $u^{0}$ with respect to $\tau$. Also note that the above solutions were obtained for an arbitrary function $f(r)$ without invoking any of its asymptotic properties or its roots, hence these hold in the case of a de-sitter spacetime as well.

Far away from the black hole, at $r\to\infty$, the spacetime is flat and one has $f(r)\to1$, the components $u^{1}$ and $u^{0}$ become,
\begin{eqnarray}
u^{1}&=&\pm\sqrt{{\left(|a|r + h\right)}^{2}-1}\label{2.infinity1}\\
u^{0}&=&|a|r + h\label{2.infinity0}
\end{eqnarray}
the solutions to which, result in the familiar Rindler trajectory with $r=\left(\cosh(|a|\tau) - h\right)/|a|$ and $t=\sinh(|a|\tau)/|a|$. Thus, it is observed that the role of the constant $h$ is to shift the Rindler hyperbole along the radial direction. Further, one can note that $dr/dt=u^{1}/u^{0}\to\pm 1$ at $r\to \infty$, as expected for an uniformly accelerated trajectory at $\tau\to\pm\infty$.
For the Schwarzschild metric, $f(r)=1-r_{s}/r$, obtaining the exact analytic form of the solutions is non trivial and one needs to deal with the Elliptic integrals of the first kind F and the of the third kind $\Pi$. To understand the general features of the solution, we first analyse the solutions for the black hole of the type, $f(r)=1-s^{2}/r^{2}$, wherein Eq.(\ref{2.result1}) is straight forwardly integrable to obtain exact analytic solution of $r(\tau)$.

\subsection{For $f(r)=1-\frac{s^2}{r^{2}}$}\label{sec2.1}

Consider the metric of the form in Eq.(\ref{2.metric}) with $f(r)=(1-s^{2}/r^{2})$,
\begin{equation}
{ds}^{2} = (1-\frac{s^{2}}{r^{2}}){dt}^{2}-{(1-\frac{s^{2}}{r^{2}})}^{-1}{dr}^{2}-r^{2}{d\theta}^{2}-r^{2}{\sin}^{2}\theta {d\phi}^{2}
\label{2.1.metric}
\end{equation}
The metric asymptotes to a flat metric at the boundary $r\to\infty$ and has a horizon at $r=+s$.

\begin{figure}[h]
\begin{subfigure}{.5\textwidth}
\includegraphics[width=7.5cm,height=5.5cm]{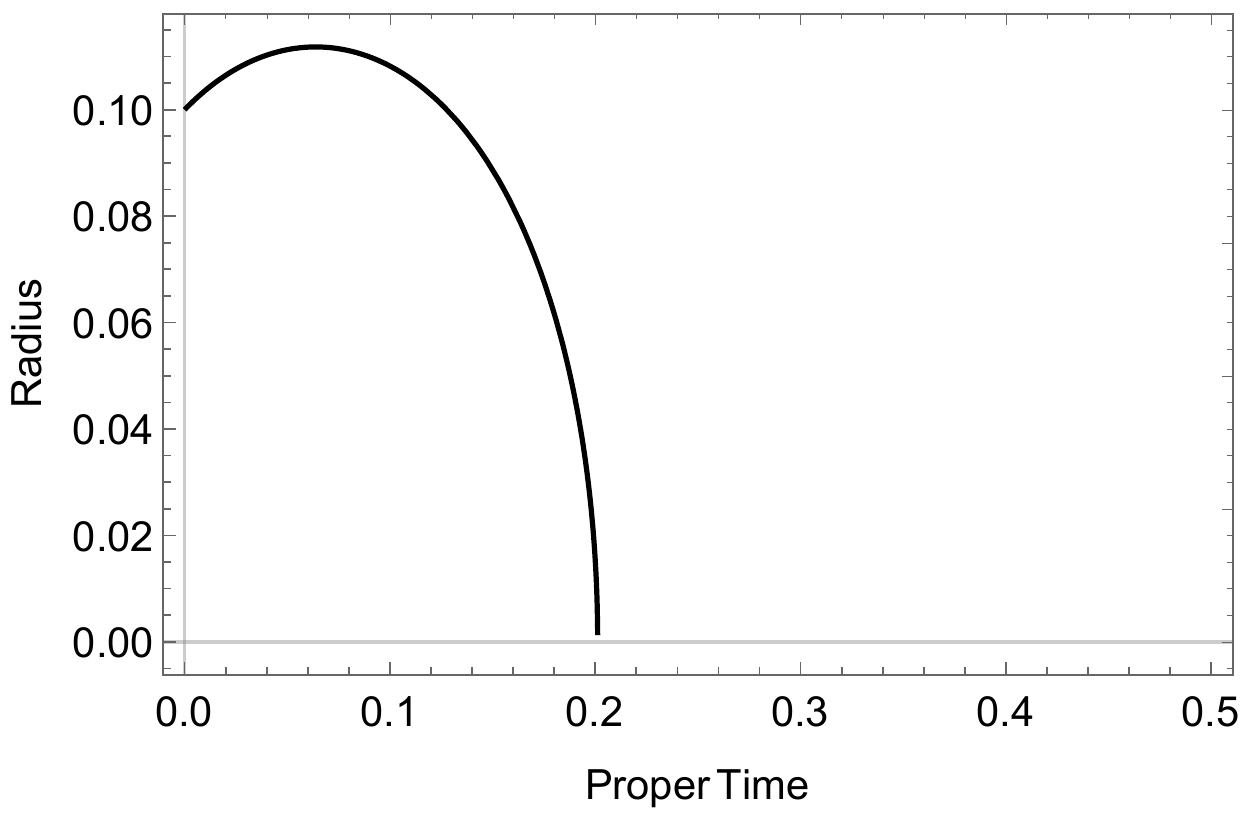}
\caption{$r_{+}(\tau)$}
\label{square-max}
\end{subfigure}
\begin{subfigure}{.5\textwidth}
\includegraphics[width=7.5cm,height=5.5cm]{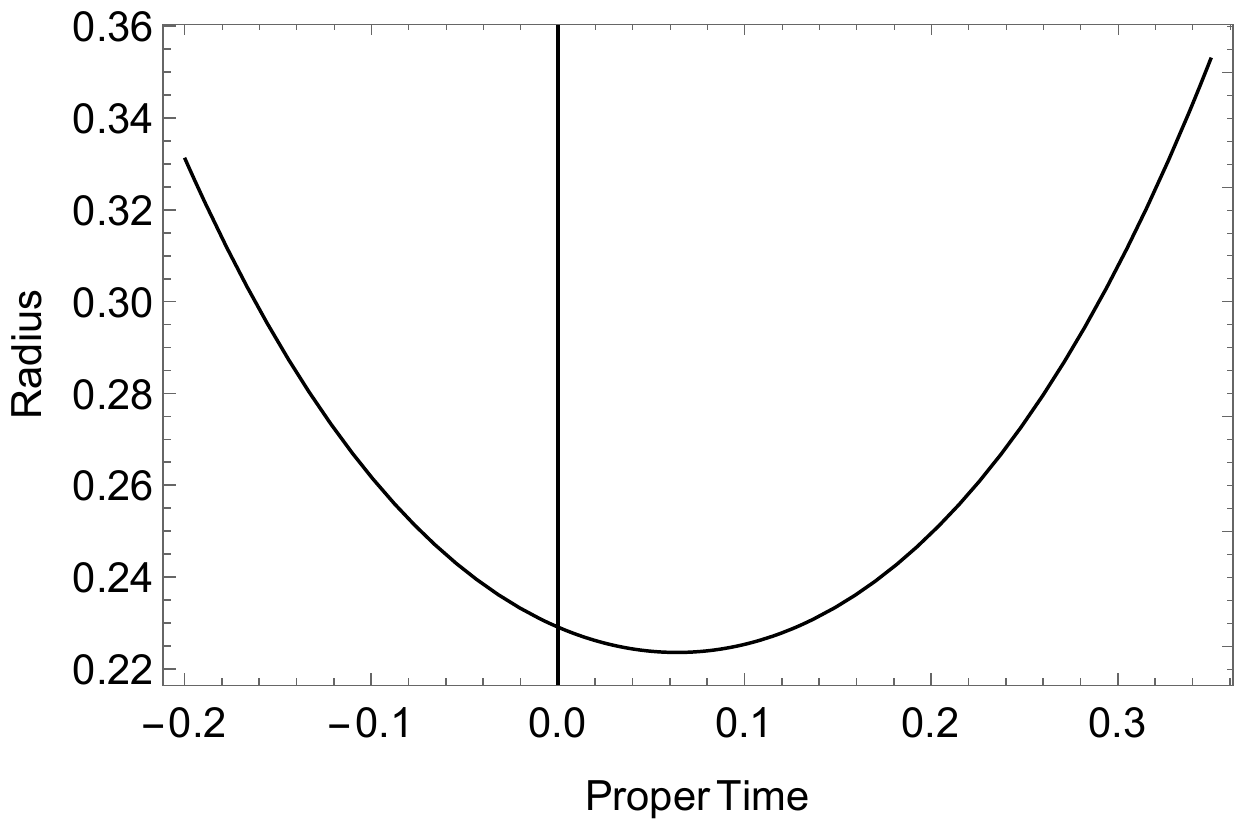}
\caption{$r_{-}(\tau)$}
\label{square-min}
\end{subfigure}
\caption{Radial trajectories $r_{+}(\tau)$ and $r_{-}(\tau)$ with $s=0.1$ and $|a|=4$. The maximum of $r_{+}$ is at $r_{max}=0.11$ and the minimum of $r_{-}(\tau)$ is at $r_{min}(\tau)=0.22$.}
\label{fig1}
\end{figure}

As in the general case, consider an observer travelling along the radial direction towards the black hole, with the initial conditions, $\tau\to -\infty$, $r>>s$ and $(dr/dt)|_{r\to -\infty}=-1$ with fixed angular coordinates. Further let the observer be constrained to be linearly uniformly accelerated such that the $3-$acceleration vector is directed radially outward. One can then expect the coordinate velocity to gradually decrease due to the outward acceleration. The gravity of the black hole of course acts in the opposite direction and the actual motion is the result of the choice of values of the parameters $|a|$ and $s$, which shall not prescribe at this junction. Without any loss of generality, consider then a trajectory which stops at some minimum value of $r$ and returns back to infinity at $\tau\to+\infty$. Such a trajectory is LUA trajectory for the metric in Eq.(\ref{2.1.metric}). Substituting $f(r)=(1-s^{2}/r^{2})$ in Eq.(\ref{2.result1}) and (\ref{2.result0}), the radial velocity is
\begin{equation}
u^{1}=-\sqrt{\left(|a|r+h\right)^2-1+\frac{s^{2}}{r^{2}}}
\label{2.1.velocity1}
\end{equation}
Since we choose the observer to be initially moving towards the black hole, the initial radial component of velocity will be negative. The time-component of four velocity is
\begin{equation}
u^{0}={(1-\frac{s^{2}}{r^{2}})}^{-1}\left(|a|r+h\right)
\label{2.1.velocity0}
\end{equation}
with a positive root, since $t$ increases as $\tau$ increases.
For the the special case $h=0$, the equation for the radial velocity can be readily integrated.
Integrating Eq.(\ref{2.1.velocity1}) with $h=0$, we get $\tau$ as a function of $r$,
\begin{equation}
\tau=\;-\frac{\log|\:(1-2{|a|}^{2}r^{2}-2|a|\sqrt{s^{2}-r^{2}+{|a|}^{2}r^{4}})\:|}{2|a|}
\label{2.1.proper-time}
\end{equation}
Inverting the function one gets two solutions for $r(\tau)$ corresponding to the two values of modulus inside the log,
\begin{eqnarray}
r_{+}(\tau)&=&\frac{1}{2|a|}\sqrt{\frac{4{|a|}^{2}s^{2}-(1-\exp(-2|a|\tau))^{2}}{\exp(-2|a|\tau)}}\label{2.1.trajectory1}\\
r_{-}(\tau)&=&\frac{1}{2|a|}\sqrt{\frac{(1+\exp(-2|a|\tau))^{2}-4{|a|}^{2}s^{2}}{\exp(-2|a|\tau)}}\label{2.1.trajectory2}
\end{eqnarray}
where the solution $r_{+}(\tau)$ corresponds to $+(1-2{|a|}^{2}r^{2}-2|a|\sqrt{s^{2}-r^{2}+{|a|}^{2}r^{4}})$ and $r_{-}(\tau)$ corresponds to $-(1-2{|a|}^{2}r^{2}-2|a|\sqrt{s^{2}-r^{2}+{|a|}^{2}r^{4}})$.
These solutions are plotted for a particular value of $|a|$ and $s$ as shown in figure(\ref{fig1}).

The trajectory given by $r_{+}(\tau)$ describes an observer travelling outward starting from some fixed radius $r$ and reaching a maximum point and then falling back into the horizon. The other solution $r_{-}(\tau)$ in Eq.(\ref{2.1.trajectory2}) describes the return trajectory with the turning point at
\begin{eqnarray}
r_{min}&=&\frac{\sqrt{1+\sqrt{1-4{|a|}^{2}s^{2}}}}{\sqrt{2}|a|}
\label{2.1.munimum}
\end{eqnarray}
For $r_{min}$ to be real, ${|a|}^{2}$ has to be less than or equal to $1/4s^{2}$.
At the bound ${|a|}^{2}=1/4s^{2}$, $r=\sqrt{2}s$, is the lowest point from which an observer can turn back. For any value of ${|a|}^{2}>1/4s^{2}$, there is no minimum and the observer falls into the black hole. We shall comment more on these bounds for the Schwarzschild case.
A similar analysis can be done for LUA observer in a metric with $f(r)$ of the form $(1-s^{n}/r^{n})$, where n is a positive integer. In such a metric, the LUA trajectory has two real positive extrema, (a) $r_{max}$ corresponding to the trajectory which initially moves away from the black hole and then turns back and falls into the horizon and (b) $r_{min}$ for the trajectory which initially falls towards the black hole and then returns back to infinity.
This can be demonstrated by applying the Descartes' rule of sign to the radial component of velocity in Eq.(\ref{2.result1}) with $f(r)=(1-s^{n}/r^{n})$ and $h=0$. We have for the radial component of the velocity
\begin{eqnarray}
u^{1}&=&\pm\sqrt{{|a|}^{2}r^{2}-1+\frac{s^{n}}{r^{n}}}
\label{2.1.n}
\end{eqnarray}
Roots of $u^{1}$ are same as the roots of the polynomial ${|a|}^{2}r^{2+n}-r^{n}+s^{n}$, which are the extrema of the LUA trajectory $r(\tau)$. This polynomial changes sign twice and hence it can have either two real positive roots or no real positive root. The two real positive roots correspond to the maximum $r_{max}$ and a minimum $r_{min}$. The bound on the parameter $|a|$ in terms of $s$ corresponds to the case when the maximum and the minimum coincide, that is, $r_{max}=r_{min}$. Thus in general, one can expect a bound to exist for the general metric with $f(r)=(1-s^{n}/r^{n})$ too.
In the next section, we analyse the LUA trajectory for the Schwarzschild case and further investigate the bounds even when $h\neq 0$.

\section{LUA observer in Schwarzschild metric}\label{sec3}

\begin{figure}[h]
\begin{subfigure}{0.5\textwidth}
\includegraphics[width=7.5cm,height=5.5cm]{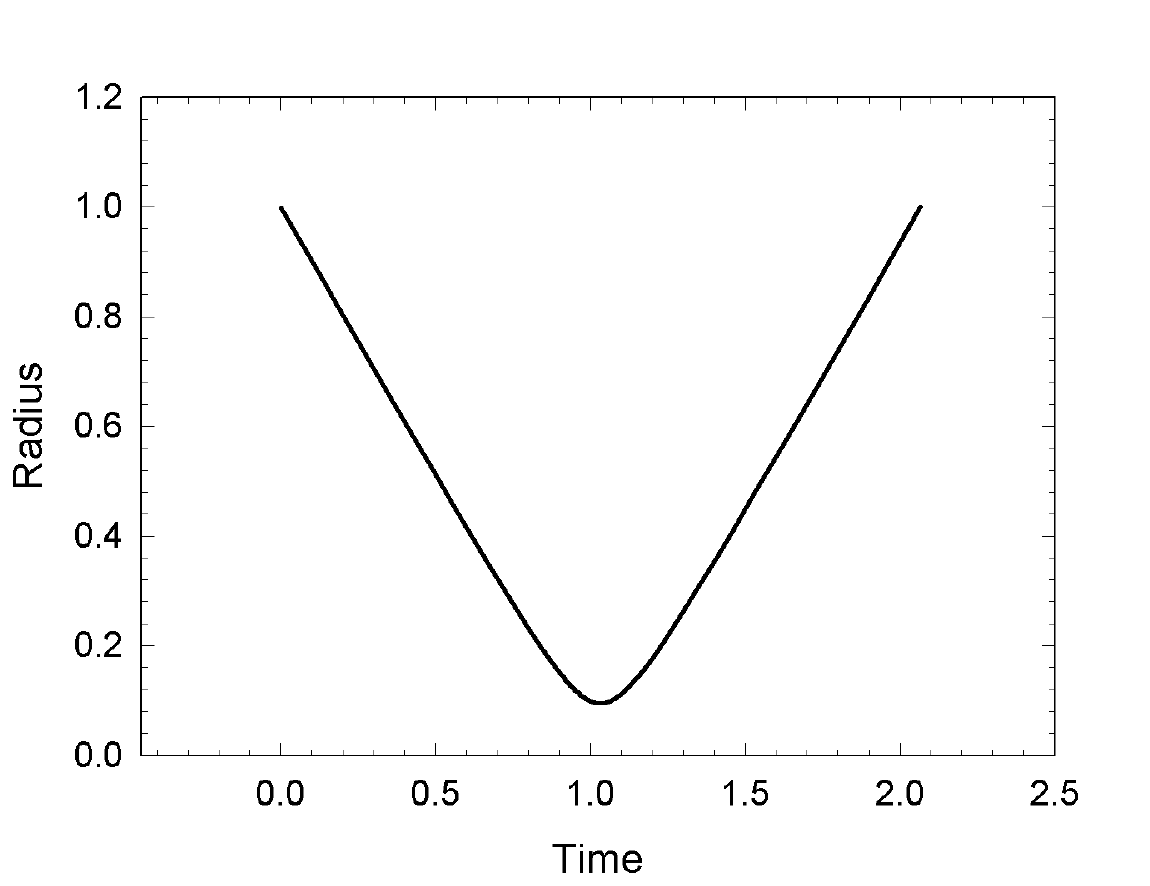}
\caption{}
\label{Sch-min}
\end{subfigure}
\begin{subfigure}{0.5\textwidth}
\includegraphics[width=7.5cm,height=5.5cm]{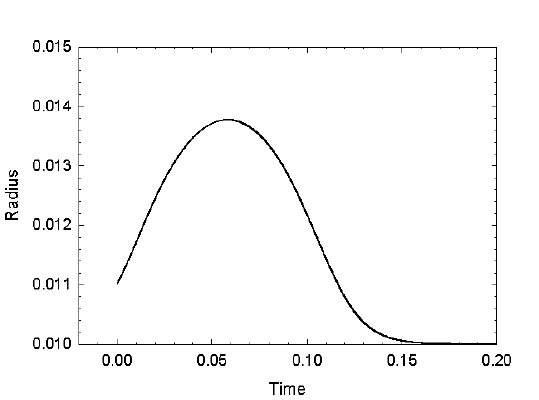}
\caption{}
\label{Sch-max}
\end{subfigure}
\caption{r-t trajectories of LUA observer (a) with initial position $r>r_{min}$, $|a|=10$, $r_{s}=0.01$ and turning point at $r_{min}=0.094564$, (b) with initial position $r<r_{max}$, $|a|=38$, $r_{s}=0.01$ and turning point at $r_{max}=0.0137713$.}
\label{fig2}
\end{figure}

For the Schwarzschild metric,
\begin{equation}
{ds}^{2} = (1-\frac{r_{s}}{r}) {dt}^{2}-{(1-\frac{r_{s}}{r})}^{-1}{dr}^{2}-r^{2}{d\theta}^{2}-r^{2}{\sin}^{2}\theta {d\phi}^{2}
\label{3.Sc-metric}
\end{equation}
consider a LUA observer travelling along a radial path with $\theta=$ constant, $\phi=$ constant initially directed towards the Schwarzschild black hole starting from some initial radius $r_i >> r_{s}$ and having uniform positive acceleration directed away from the black hole. The components of velocity vector for the LUA observer are given by Eq.(\ref{2.result1}) and (\ref{2.result0}) with $f(r) = (1-r_{s}/r)$.
\begin{eqnarray}
u^{1}&=&\dfrac{dr}{d\tau}=\pm\sqrt{\left(|a|r+h\right)^2-1+\frac{r_{s}}{r}}\label{3.velocity1}\\
u^{0}&=&\dfrac{dt}{d\tau}={(1-\frac{r_{s}}{r})}^{-1}\left(|a|r+h\right)\label{3.velocity0}
\end{eqnarray}
Consider first the special case, $h=0$, wherein the components of velocity vector simplify to,
\begin{eqnarray}
u^{1}=&\dfrac{dr}{d\tau}=&\pm\sqrt{{|a|}^{2}r^{2}-1+\frac{r_{s}}{r}}\label{3.c-velocity1}\\
u^{0}=&\dfrac{dt}{d\tau}=&{(1-\frac{r_{s}}{r})}^{-1}|a|r\label{3.c-velocity0}
\end{eqnarray}
As mentioned in the earlier section, for the case $h=0$, taking the limit $r_s \rightarrow 0$ in the above equations corresponds to the Rindler trajectory in the usual form. The equation of motion $r(t)$ is obtained by taking $u^{1}/u^{0}$ to get,
\begin{eqnarray}
\dfrac{dr}{dt}&=&\pm(1-\frac{r_{s}}{r})\frac{\sqrt{{|a|}^{2}r^{2}-1+\frac{r_{s}}{r}}}{|a|r}
\label{3.c-motion}
\end{eqnarray}
The solutions for the equation of motion can be written in terms of the Elliptic integrals of the first kind $F$ and the of the third kind $\Pi$. Solving the differential equation numerically for different values of acceleration $|a|$ and $r_{s}$, the radius $r$ is plotted as a function of time coordinate $t$ (Figure \ref{fig2}).

From the figure one can check that the trajectory starts from a initial radius, falls towards the black hole and then returns back to a large value of $r$, from the turning point at radius $r_{min} > r_{s}$ as expected. To determine the extrema of the trajectory, we take $dr/dt=0$, which leads to a cubic polynomial ${|a|}^{2}r^{3}-r+r_{s}=0$. As per the Descartes' rule of sign, two roots of this polynomial are positive real or none of them are. We shall now investigate the constraints under which, the minimum and maximum exist. The roots of ${|a|}^{2}r^{3}-r+r_{s}=0$ are  
\begin{eqnarray}
r_{min} =& r_{1} =&  \frac{{\left(\frac{2}{3}\right)}^{1/3}}{{\left(-A_{0}+B_{0}\right)}^{1/3}}+\frac{{\left(-A_{0}+B_{0}\right)}^{1/3}}{{2}^{1/3}\, 3^{2/3}\,{|a|}^{2}}\\
& r_{2} =& -\frac{(1+i\sqrt{3})}{2^{2/3}\,3^{1/3}\,{\left(-A_{0}+B_{0}\right)}^{1/3}}-\frac{(1-i\sqrt{3}) {\left(-A_{0}+B_{0}\right)}^{1/3}}{{2}^{4/3}\,3^{2/3}\,{|a|}^{2}}\\
r_{max} =& r_{3} =& -\frac{(1-i\sqrt{3})}{2^{2/3}\,3^{1/3}\,{\left(-A_{0}+B_{0}\right)}^{1/3}}-\frac{(1+i\sqrt{3}) {\left(-A_{0}+B_{0}\right)}^{1/3}}{{2}^{4/3}\,3^{2/3}\,{|a|}^{2}}
\end{eqnarray}
where, $A_{0}=9{|a|}^{4}r_{s}$ and $ B_{0}=\sqrt{3}{|a|}^{3}\sqrt{27{|a|}^{2}{r_{s}}^{2}-4}$.
For ${|a|}^{2}>(4/27{r_{s}}^{2})$, $(-A_{0}+B_{0})$ is real and thus $r_{2}$ and $r_{3}$ become complex valued and $r_{1}$ is negative. Thus for values of acceleration ${|a|}^{2}>(4/27{r_{s}}^{2})$, there is no turning point and the LUA observer falls into the horizon.

For ${|a|}^{2}\leq (4/27{r_{s}}^{2})$, the term $(-A_{0}+B_{0})$ is complex and can be written as $(-A_{0}+B_{0})=(-A_{0}+iB^{'}_{0})=C_{0}\exp\left(i(\pi-\eta)\right)$, where, $B^{'}_{0}=\sqrt{3}{|a|}^{3}\sqrt{4-27{|a|}^{2}{r_{s}}^{2}}$, $C_{0}=\sqrt{{A_{0}}^{2}+{B^{'}_{0}}^{2}}$ and $\eta={\tan}^{-1}(B^{'}_{0}/A_{0})$. Using this, the expressions $r_{1}$, $r_{2}$ and $r_{3}$ simplify to be
\begin{eqnarray}
r_{min} =& r_{1} =& \frac{\left(\cos(\eta/3)+\sqrt{3}\sin(\eta/3)\right)}{\sqrt{3}|a|}\label{3.c-extremum1}\\
& r_{2} =&\frac{-4\cos(\eta/3)}{2\sqrt{3}|a|}\label{3.c-extremum2}\\
r_{max} =& r_{3} =&  \frac{\left(\cos(\eta/3)-\sqrt{3}\sin(\eta/3)\right)}{\sqrt{3}|a|}\label{3.c-extremum3}
\end{eqnarray}
From the definition of $\eta$, it can take values from $0$ to $\pi/2$. For this range of $\eta$, both $\cos(\eta/3)$ and $\sin(\eta/3)$ are positive and hence $r_{2}$ is negative and thus not physical, while $r_{1}$ is a point of minimum and $r_{3}$ is a point of maximum. Here $r_{3}$, the point of maximum refers to LUA trajectories which starts from radius $r_i < r_3$ with an initial velocity in the outward direction, reach a maximum value of $r_{3}$ and then turn and fall back into the black hole, similar to the trajectory in Figure (\ref{Sch-max}).

A LUA trajectory having acceleration ${|a|}^{2}<(4/27{r_{s}}^{2})$ is shown in figure(\ref{Sch-min}), with the point of minimum given by $r_{1}$ i.e. Eq.(\ref{3.c-extremum1}). Whereas the observer with acceleration ${|a|}^{2}>(4/27{r_{s}}^{2})$ does not have a minimum and falls into the black hole.

As the magnitude of acceleration ${|a|}^{2}$ approaches the saturation value of the bound $(4/27{r_{s}}^{2})$ from below, both the extrema, that is, the minimum $r_{1}$ and the maximum $r_{3}$ come closer. At the exact value, ${|a|}^{2}=(4/27{r_{s}}^{2})$, $B_{0}^{'}$ vanishes and hence $\sin (\eta/3)$ is zero, thus making the two extrema equal at radius
\begin{equation}
r_b=\frac{1}{\sqrt{3}|a|}=\frac{3r_{s}}{2}
\label{3.c-extremum at bound}
\end{equation}
which is greater than the Schwarzschild radius $r_s$.

For the general case, $h\neq 0$, the equation of motion for the LUA trajectory from Eqs.(\ref{3.velocity1}) and (\ref{3.velocity0})  is ,
\begin{eqnarray}
\dfrac{dr}{dt}&=&\frac{u^{1}}{u^{0}}=\pm(1-\frac{r_{s}}{r})\frac{\sqrt{\left(|a|r+h\right)^2-1+\frac{r_{s}}{r}}}{\left(|a|r+h\right)}
\label{3.motion}
\end{eqnarray}
Solving this differential equation numerically, the trajectory $r$ as a function of $t$ is plotted to get the curves  similar to that of figure(\ref{fig2}).
The extrema of this curve r(t) are obtained by solving $dr/dt=0$, to get the following three roots,
\begin{eqnarray}
r_{min} =& r_{1} =&-\frac{2h}{3|a|}+\frac{P^{1/3}}{2^{1/3}\:3{|a|}^{2}}+\frac{2^{1/3}(3+h^2)}{3\:P^{1/3}}\\
& r_{2} =&-\frac{2h}{3|a|}-\frac{(1-i \sqrt{3})P^{1/3}}{{2}^{1/3}\:6{|a|}^{2}}-\frac{(1+i\sqrt{3})(3+h^{2})}{2^{2/3}\:3\:P^{1/3}}\\
r_{max} =& r_{3} =&-\frac{2h}{3|a|}-\frac{(1+i \sqrt{3})P^{1/3}}{{2}^{1/3}\:6{|a|}^{2}}-\frac{(1-i\sqrt{3})(3+h^{2})}{2^{2/3}\:3\:P^{1/3}}
\end{eqnarray}
here,
$P=-A+iB$, with, $A=(27{|a|}^{4}r_{s}+18{|a|}^{3}h-2{|a|}^{3}h^{3})$ and $B=\sqrt{4{|a|}^{6}(3+h^2)^{3}-(A)^{2}}$. Simplifying these expressions in a similar manner as in the case $h=0$, we get,
\begin{eqnarray}
r_{min} =& r_{1} =&\frac{2}{3|a|}\left[\frac{\sqrt{3+h^{2}}}{2}\left(\cos(\xi/3)+\sqrt{3}\sin(\xi/3)\right)-h\right]\label{eq-Sch-min}\\
& r_{2} =&\frac{-2}{3|a|}\left[\sqrt{3+h^{2}}\cos(\xi/3)+h\right]\label{eq-Sch-2}\\
r_{max} =& r_{3} =&\frac{2}{3|a|}\left[\frac{\sqrt{3+h^{2}}}{2}\left(\cos(\xi/3)-\sqrt{3}\sin(\xi/3)\right)-h\right]\label{eq-Sch-max}
\end{eqnarray}
where, $\xi=\tan^{-1}(B/A)$. For $r_{1}$, $r_{2}$ and $r_{3}$ to be real, $B$ needs to be real, that is, $4{|a|}^{6}(3+h^2)^{3}-(A)^{2}$ should be greater than or equal to zero, and thus $\xi$ will be in the range $0<\xi<\pi$. It is then clear from the equations (\ref{eq-Sch-min}), (\ref{eq-Sch-2}), (\ref{eq-Sch-max}) that, $r_{2}$ will always be negative since $\cos(\xi/3)$ is positive, and thus it does not represent a physical solution. The root $r_{1}$ gives us the minimum of the trajectory that we are interested in.

\begin{figure}[h]
\centering
\includegraphics[width=7.5cm,height=5.5cm]{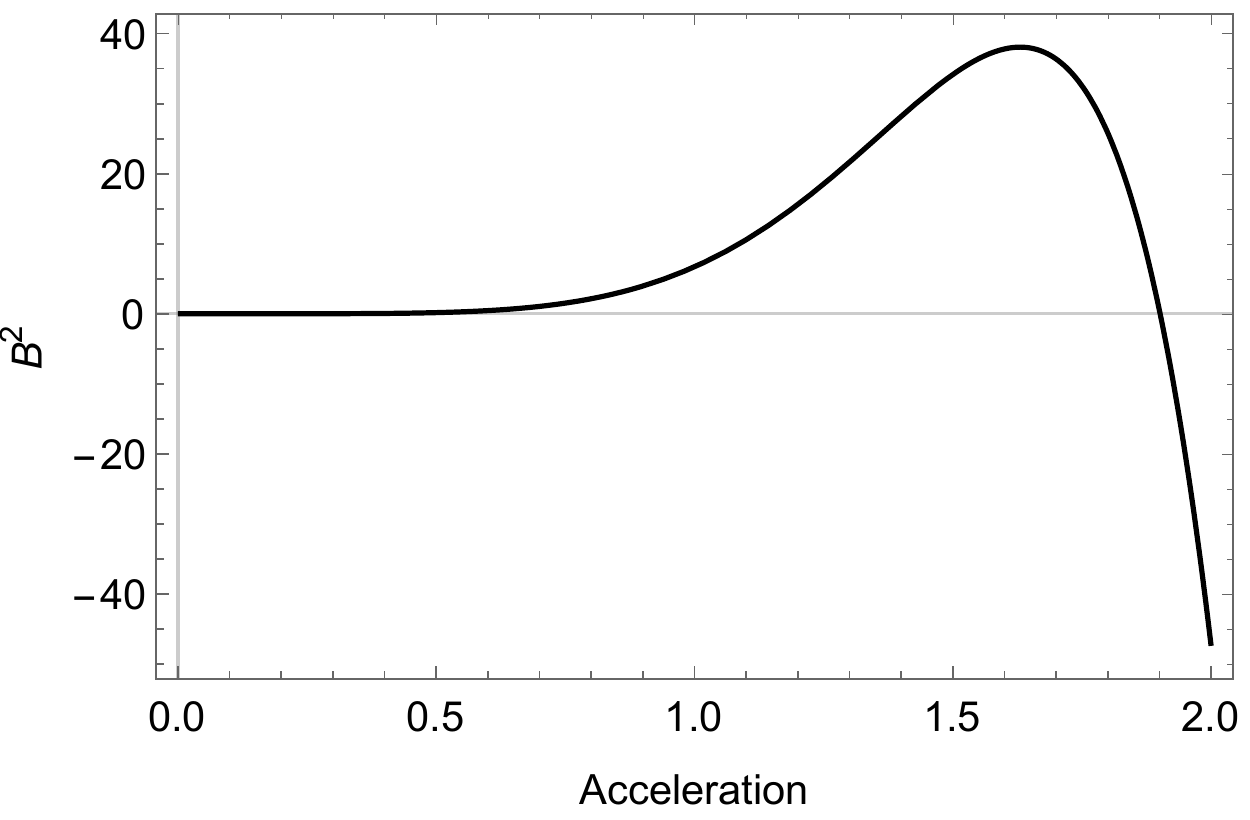}\caption{Plot of $B^2 = 4{|a|}^{6}(3+h^2)^{3}-(27{|a|}^{4}r_{s}+18{|a|}^{3}h-2{|a|}^{3}h^{3})^{2}$ as a function of $|a|$ with $h=0.8$ and $r_{s}=0.01$. The non negative zeros of the function are $a_{0}=0$ and $a_{2}=1.90131$.} 
\label{B^2}
\end{figure}

To find the range of acceleration for which, $B$ is real, i.e $B^{2}\geq 0$, we first obtain the values of acceleration $|a|$ which give the zeros of function $B$, that is imposing $B=0$, the acceleration $|a|$ is obtained in terms of $h$ and $r_{s}$. The real solutions are,
\begin{eqnarray}
a_{1}&=&\frac{2(-9h+h^3-\sqrt{(3+h^2)^3})}{27r_{s}} \nonumber \\
a_{2}&=&\frac{2(-9h+h^3+\sqrt{(3+h^2)^3})}{27r_{s}}=\mathcal{B}(M,h)
\label{3.accsol}
\end{eqnarray}
One can check that, $a_{1}$ is a negative value for whole range of $h$ and thus not a physical solution and $B^{2}\geq 0$ for the range $0\leq |a|\leq a_{2}=\mathcal{B}(M,h)$. A plot of $B^{2}$ as a function of $|a|$ for a particular value of $h$ and $r_s$ is shown in figure(\ref{B^2}).

\begin{figure}[h]
\begin{subfigure}{0.5\textwidth}
\includegraphics[width=7.5cm, height=5.5cm]{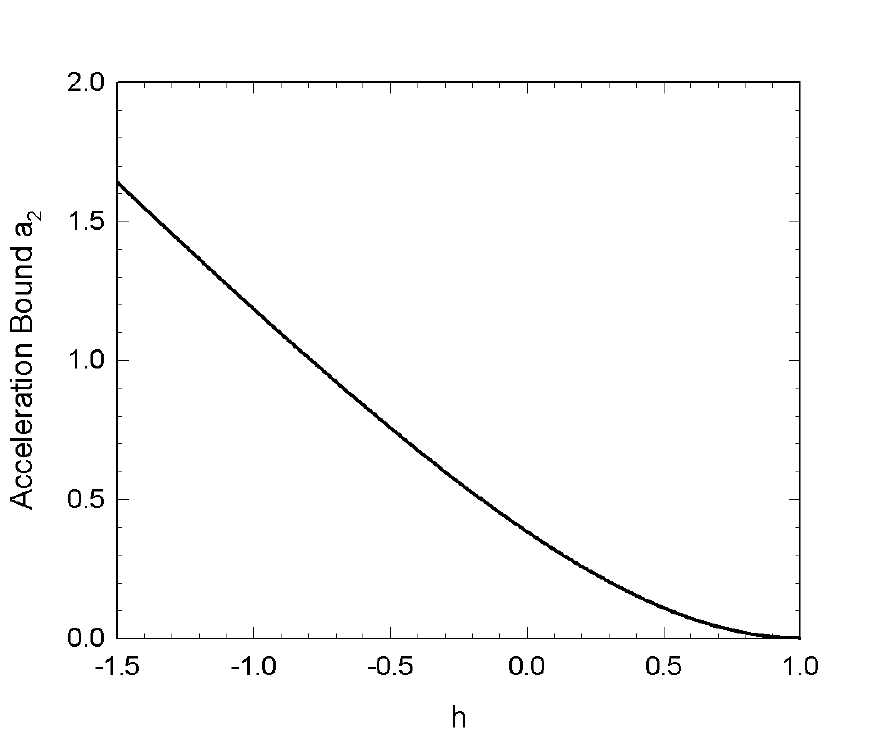} 
\caption{}
\label{acc-bound}
\end{subfigure}
\begin{subfigure}{0.5\textwidth}
\includegraphics[width=7.5cm, height=5.5cm]{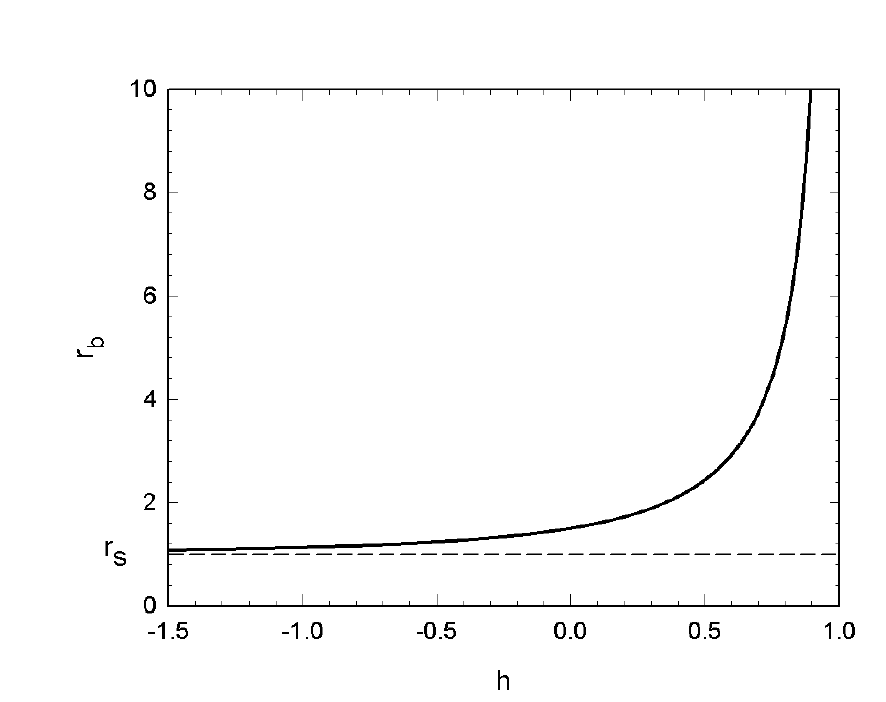}
\caption{}
\label{radius-bound}
\end{subfigure}
\caption{$a_2$ and $r_b$ as a function of $h$ for $r_s=1$. The distance of closest approach $r_{b}$ tends to $r_s$ as $h\to -\infty$.}
\label{bounds}
\end{figure}

\begin{figure}[h]
\centering
\includegraphics[width=7.5cm,height=5.5cm]{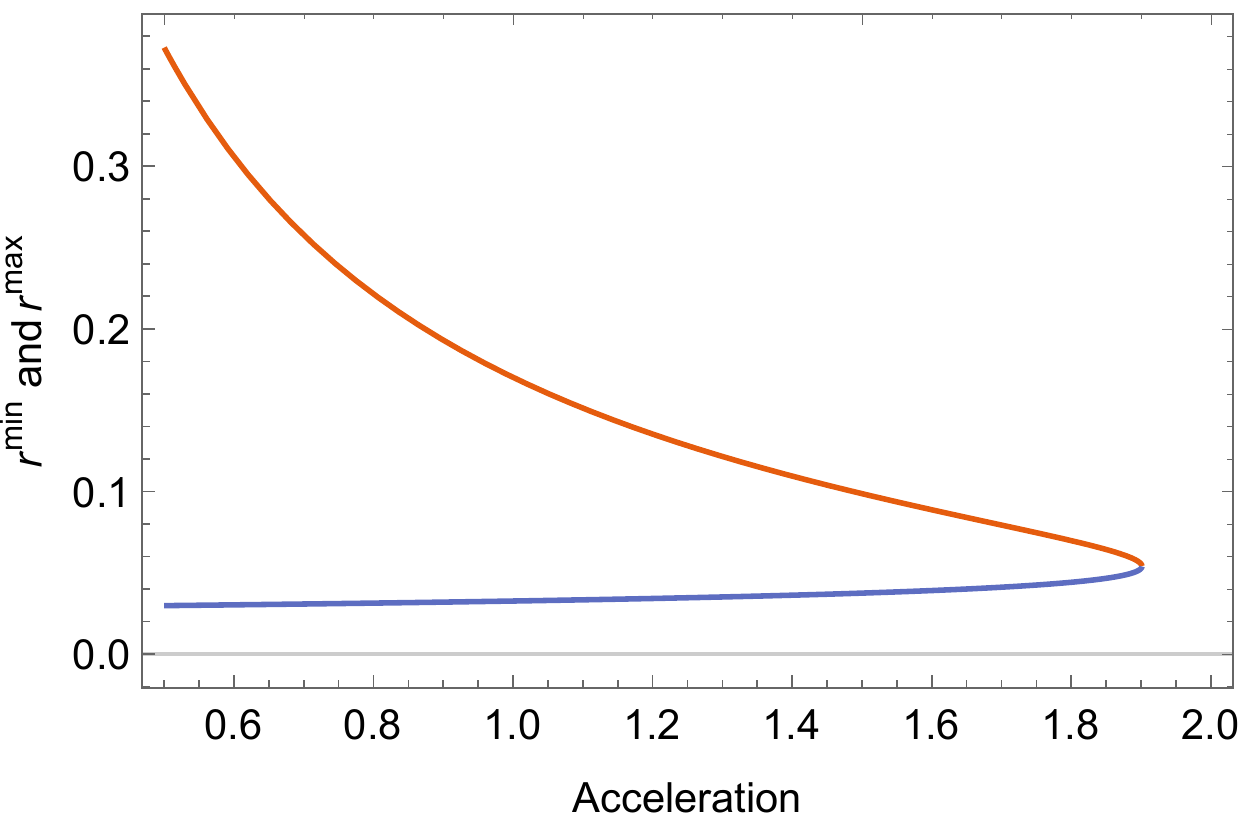}
\caption{$r_{1} = r_{min}$ and $r_{3} = r_{max}$ as a function of $|a|$ for $h=0.8$ and $r_{s}=0.01$. The extrema $r_{1}$ and $r_{3}$ merge at bound $|a|=a_{2}=1.90131$}
\label{r1r3}
\end{figure}

 At the bound $|a|=a_{2}$, since $B^{2}=0$, $\sin(\xi/3)$ is zero and the two extrema $r_{1}$ and $r_{3}$ equal to
\begin{eqnarray}
r_{b}&=& \frac{2}{3|a|}\left(\frac{\sqrt{3+h^2}}{2}-h\right)\nonumber\\
&=&\frac{9r_{s}\left(\sqrt{3+h^2}-2h\right)}{2\left({-9h+h^3+\sqrt{(3+h)^3}}\right)}
\label{3.bound}
\end{eqnarray}
Eq.(\ref{3.bound}) gives the distance of closest approach $r_b$ beyond which the LUA observer cannot have a turning point, which is positive only for the values $h<1$. The bound on acceleration $a_{2}$ and the distance of closest approach $r_{b}$ are plotted as a function of $h$ for $r_s=1$ in figure(\ref{bounds}). The acceleration $a_{2}$ is zero at $h=1$ and thus $r_{b}$ has an indeterminate form $0/0$ at $h=1$.
The minimum $r_{1}$ and maximum $r_{3}$ are also plotted in figure(\ref{r1r3}) below for a value of $h<1$ and $r_{s}=0.01$. It can be seen that the extrema $r_{1}$ and $r_{3}$ merge at the bound $|a|=a_{2}$.

Thus to have a radial LUA trajectory with a turning point in Schwarzschild metric, it should have the initial conditions with the constant $h<1$ and the magnitude of acceleration in range $0<|a|\leq a_{2}$. The bound on the value of acceleration is then given by Eq.(\ref{3.accsol}) and the corresponding distance of closest approach is given by Eq.(\ref{3.bound}) for the particular value of $h$. For a different choice of constants, the bounds are derived in Appendix A.

Similar to the Rindler trajectory in flat spacetime, the parameter $h$ plays the role of shifting the LUA trajectory in the Schwarzschild case. For $h = 0$, the LUA trajectory far away from the black hole, $r \rightarrow \infty$, matches with the Rindler hyperbola in flat spacetime at asymptotic infinity, with $r=0$ and $t=0$ being the bifurcation point of the Killing horizon. Then, decreasing or increasing $h$ amounts to shifting the trajectory to the right or left in the $(t,r)$ plane at flat asymptotic infinity. Of-course, at finite $r$ or close to the black hole, the particular solution of the LUA trajectory, for a chosen $h$, may change non-linearly with $h$. However, for the family of Rindler curves corresponding to the Rindler quadrant (bounded by the casual past of the future asymptotic point and the casual future of the past asymptotic point and future null infinity and past null infinity) greater the shift towards the black hole; the trajectories with higher acceleration from the family of Rindler curves, will be affected more by the curvature of the black hole since they are relatively closer to the Rindler horizon. This picture is consistent with the results of Eq.(\ref{3.accsol}) and Eq.(\ref{3.bound}). From Figure (\ref{acc-bound}) for the bound on acceleration $a_2$, one can see that the bound increases with the decrease in value of $h$. That is, as the Rindler quadrant is shifted away from the black hole, the limit on trajectories with higher acceleration which can return back to infinity, increases as $h$ decreases and vice-versa. 

To identify the family of radial LUA trajectories bounded by a particular Rindler quadrant, one needs to know the future and past asymptotic points of the radial LUA trajectories. However, the solution for the trajectories in terms of the elliptic functions present a technical difficulty in obtaining a value for the asymptotes. This is due to the fact that the incomplete elliptic integral of the third kind one encounters in the solution is not well defined at $r \rightarrow \infty$ in the present case and hence one needs to resort to other methods which we shall describe in a future publication \cite{kajol}. One can though expect the asymptotic points to depend on both the values of $|a|$ and $h$ in general. Interestingly this would imply that the Rindler quadrant would be different for radial LUA trajectories having different $|a|$ and $h$.

\section{For the De-Sitter spacetime:}\label{sec4}

It is known that $4D$ De-sitter spacetime can be embedded in $5D$ Minkowskian space $({{\Gamma}^{a}}_{bc}=0)$ as a hyperboloid with constant curvature $\mathcal{R}$. The magnitude of $4-$acceleration of a radial trajectory in $4D$ De-sitter spacetime is related to the magnitude of the $5-$acceleration of the same trajectory in the embedding flat space as\cite{deser},
\begin{equation}
{|a|}^{2}_{5D}={|a|}^{2}+\frac{1}{R^{2}}
\label{4.acc relation}
\end{equation}
where, ${|a|}^{2}=-g_{ij}a^{i}a^{j}$ is the magnitude of $4-$acceleration. Hence, a constant $4-$acceleration implies a uniform acceleration in $5D$ as well. Now, a linearly uniformly accelerated trajectory defined in $5D$ is just the usual hyperbolic Rindler trajectory in $5D$. It would be interesting to ask whether the linearity in embedded $4D$ De-sitter spacetime in terms of ${[w^{i}-{|a|}^{2}u^{i}]}_{4D}=0$ equations also implies the corresponding linearity equation ${[W^{b}-{|a|}^{2}U^{b}]}_{5D}=0$ in the embedding $5D$ space. In this section, we investigate the one to one correspondence between the LUA trajectory in $5D$ and the LUA trajectory in $4D$ space.

The De-Sitter metric in the spherical coordinates $x^{i}\equiv(t,r,\theta,\phi)$ is
\begin{equation}
{ds}^{2} = (1-\frac{r^{2}}{R^{2}}){dt}^{2}-{(1-\frac{r^{2}}{R^{2}})}^{-1}{dr}^{2}-r^{2}{d\theta}^{2}-r^{2}{\sin}^{2}\theta {d\phi}^{2}
\label{4.metric}
\end{equation}
with $1/R^{2}=\Lambda/3$, where $\Lambda$ is the Cosmological constant.

To get the solution $r(\tau)$ and $t(\tau)$ for the trajectory of a LUA observer moving along radial direction ($\theta=constant$ and $\phi=constant$) in De-Sitter metric, we have used the results of section \ref{sec2}. Even if $f(r)=(1-r^{2}/R^{2})$ for De-Sitter metric does not satisfy the condition, $f(r)\to1$ as $r\to\infty$, the results of section \ref{sec2} can be used for the LUA observer in this metric, as this property of $f(r)$ is not used anywhere in the derivation of the results Eq.(\ref{2.result1}) and Eq.(\ref{2.result0}).
By substituting $f(r)=(1-r^{2}/R^{2})$ in equations (\ref{2.result1}) and (\ref{2.result0}) the components of velocity vector become,
\begin{eqnarray}
u^{1}&=&\dfrac{dr}{d\tau}=\sqrt{\left(|a|r+h\right)^2-1+\frac{r^2}{R^2}}\label{4.velocity1}\\
u^{0}&=&\dfrac{dt}{d\tau}={(1-\frac{r^2}{R^2})}^{-1}\left(|a|r+h\right)\label{4.velocity0}
\end{eqnarray}
Solving the differential equation (\ref{4.velocity1}), $r(\tau)$ is obtained as,
\begin{eqnarray}
r(\tau)&=&\frac{A \exp(\chi\tau) -B \exp(-\chi\tau) + |a|R\sqrt{1+4AB}}{\chi}
\label{4.r trajectory}
\end{eqnarray}
where, $\chi=\sqrt{(1/R^{2})+{|a|}^{2}}$, $A = e^{(\chi k)}/2R^{2}\chi$ and $B =\left(h^{2}-\chi^2 R^{2}\right)e^{(-\chi k)}/2\,\chi$ with $k$ as the constant of integration.

Substituting this $r$ in Eq.(\ref{4.velocity0}) and integrating it with respect to $\tau$, the solution $t(\tau)$ is obtained as,
\begin{eqnarray}
t(\tau)&=&R\left[\tanh^{-1}\left(\frac{\chi^2 R^{3}-|a|hR^2+e^{\chi(k+\tau)}}{\chi R^{2}\quad(|a|R-h)}\right)\right]\nonumber\\
& &-R\left[\tanh^{-1}\left(\frac{\chi^2 R^{3}+|a|hR^2-e^{\chi(k+\tau)}}{\chi R^{2}\quad(|a|R+h)}\right)\right]
\label{4.t trajectory}
\end{eqnarray}

The Eq.(\ref{4.r trajectory}) and Eq.(\ref{4.t trajectory}) thus give the trajectory of a LUA observer in De-Sitter space.

\subsection{5D Embedding flat space}\label{sec4.1}

To get the trajectory of LUA observer in the embedding space, consider first the relations between coordinates $z^{a}$ in $5D$ emmbedding space and coordinates $x^{i}\equiv(t,r,\theta,\phi)$ in $4D$ De-Sitter space\cite{deser},
\begin{equation}
z^{0}=\sqrt{R^{2}-r^{2}} \sinh(t/R)\qquad
z^{1}=\sqrt{R^{2}-r^{2}} \cosh(t/R)\qquad
z^{4}=r\cos(\theta)
\label{4.1.coordinates1}
\end{equation}
\begin{equation}
z^{2}=r\sin(\theta)\cos(\phi)\qquad
z^{3}=r\sin(\theta)\sin(\phi)\qquad
\label{4.1.coordinates2}
\end{equation}
The De-Sitter metric is then obtained through ${ds}^{2}=\eta_{ab}{dz}^{a}{dz}^{b}$ with the above substitution for $z^{a}$. For radial trajectory (choosing $\theta=0$), $z^{2}=z^{3}=0$ and $z^{4}=r$, and the components of velocity vector, ${U}^{a}=dz^{a}/d\tau$ in the $5D$ space are,
\begin{eqnarray}
U^{0}&=&\frac{\sqrt{R^{2}-r^{2}}}{R} \cosh(t/R) u^{0}-\frac{r}{\sqrt{R^{2}-r^{2}}} \sinh(t/R) u^{1}\label{4.1.velocity0}\\
U^{1}&=&\frac{\sqrt{R^{2}-r^{2}}}{R} \sinh(t/R) u^{0}-\frac{r}{\sqrt{R^{2}-r^{2}}} \cosh(t/R) u^{1}\label{4.1.velocity1}\\
U^{4}&=&u^{1}\label{4.1.velocity4}\\
U^{2}&=&U^{3}\;=\;0\label{4.1.velocity23}
\end{eqnarray}
here, $u^{0}=dt/d\tau$ and $u^{1}=dr/d\tau$ are the components of $4D$ velocity vector.

The components of acceleration vector, $({{a}^{b}})_{5D}=dU^{b}/d\tau$ in terms of the components of four velocity are,
\begin{eqnarray}
({{a}^{0}})_{5D}&=&\frac{\cosh(t/R)}{R}\Big[\sqrt{R^{2}-r^{2}}\dfrac{du^{0}}{d\tau}-\frac{2ru^{0}u^{1}}{\sqrt{R^{2}-r^{2}}}\Big] +\frac{\sinh(t/R)}{\sqrt{R^{2}-r^{2}}}\Big[1-r\dfrac{du^{1}}{d\tau}\Big]\label{4.1.acc0}\\
({{a}^{1}})_{5D}&=&\frac{\sinh(t/R)}{R}\Big[\sqrt{R^{2}-r^{2}}\dfrac{du^{0}}{d\tau}-\frac{2ru^{0}u^{1}}{\sqrt{R^{2}-r^{2}}}\Big] +\frac{\cosh(t/R)}{\sqrt{R^{2}-r^{2}}}\Big[1-r\dfrac{du^{1}}{d\tau}\Big]\label{4.1.acc1}\\
({{a}^{4}})_{5D}&=&\dfrac{du^{1}}{d\tau}\label{4.1.acc4}\\
({{a}^{2}})_{5D}&=&({{a}^{3}})_{5D}=0\label{4.1.acc23}
\end{eqnarray}

In $5D$ Minkowskian spacetime, the constraints for the trajectory to be LUA are simply $W^{b}-{{|a|}^{2}}_{5D}U^{b}=0$ or,
\begin{eqnarray}
0&=W^{0}-{{|a|}^{2}}_{5D}U^{0}&=\dfrac{d^{2}U^{0}}{d{\tau}^{2}}-{{|a|}^{2}}_{5D}U^{0}\label{4.1.linearity0}\\
0&=W^{1}-{{|a|}^{2}}_{5D}U^{1}&=\dfrac{d^{2}U^{1}}{d{\tau}^{2}}-{{|a|}^{2}}_{5D}U^{1}\label{4.1.linearity1}\\
0&=W^{4}-{{|a|}^{2}}_{5D}U^{4}&=\dfrac{d^{2}U^{4}}{d{\tau}^{2}}-{{|a|}^{2}}_{5D}U^{4}\label{4.1.linearity4}
\end{eqnarray}
where, ${{|a|}^{2}}_{5D}=-\eta_{ab}({{a}^{a}})_{5D}({{a}^{b}})_{5D}$ is the $5-$acceleration.
The other two components of $W^{b}$ are zero.
Substituting $U^{0},U^{1}$ and $U^{4}$ given by equations (\ref{4.1.velocity0}), (\ref{4.1.velocity1}) and (\ref{4.1.velocity4}) in the above equations and simplifying them using the normalization $(g_{ij}u^{i}u^{j}=1)$ and orthogonality condition ($g_{ij}u^{i}a^{j}=0$, where $a^{j}=u^{k}\nabla_{k}u^{j}$), we can relate the $5D$ quantities to the $4D$ quantities as
\begin{eqnarray}
W^{0}-{{|a|}^{2}}_{5D}U^{0}&=&\frac{\cosh(t/R)\sqrt{R^{2}-r^{2}}}{R}{[w^{0}-{|a|}^{2}u^{0}]}_{4D}\label{4.1.result0}\\
W^{1}-{{|a|}^{2}}_{5D}U^{1}&=&\frac{\sinh(t/R)\sqrt{R^{2}-r^{2}}}{R}{[w^{0}-{|a|}^{2}u^{0}]}_{4D}\label{4.1.result1}\\
W^{4}-{{|a|}^{2}}_{5D}U^{4}&=&{[w^{1}-{|a|}^{2}u^{1}]}_{4D}\label{4.1.result4}
\end{eqnarray}

From above equations, it is clear that, demanding the radial trajectory in $4D$ De-Sitter space to satisfy the linearity conditions implies the linearity in embedding $5D$ space. The trajectory of the LUA observer ($\,z^{0}(\tau),\,z^{1}(\tau),\,z^{4}(\tau)\,$) in embedding space then can be simply obtained by substituting of $r(\tau)$ and $t(\tau)$ given by Eq.(\ref{4.r trajectory}) and Eq.(\ref{4.t trajectory}) in Eq.(\ref{4.1.coordinates1}).
We have demonstrated this for a particular radial LUA trajectory in $4D$ De-Sitter space, which gives the usual $2D$ Rindler hyperbolic trajectory in the embedding $5D$ space. The trajectory with constants $h=0$ and $k=\log(\chi R^{2})/\chi$, in $4D$ De-Sitter space is,
\begin{equation}
r(\tau)= \frac{\cosh(\chi\tau)}{\chi}, \qquad t(\tau)=R\, {\tanh}^{-1}(\frac{\sinh(\chi\tau)}{|a| R}) 
\end{equation}
which gives the trajectory in embedding space as,
\begin{equation}
z^{0}=\frac{\sinh(\chi\tau)}{\chi},\qquad
z^{1}=\frac{|a| R}{\chi},\qquad
z^{4}=\frac{\cosh(\chi\tau)}{\chi}
\end{equation}
Thus the LUA trajectory in embedding $5D$ space is the usual Rindler trajectory with frame of reference translated along $z^{1}$ direction by a constant value $|a|R/\chi$, for the above choice of constants $h$ and $k$.

\section{Discussion}

The lower bound of $r_{b} >2 M$ on the distance of closest approach $r_{min}$ to the Schwarzschild black hole for a linearly uniformly accelerating observer is indeed an intriguing one. For the $h=0$ case, the LUA trajectory far away from the black hole, $r \rightarrow \infty$, matches with the Rindler hyperbola in flat spacetime at asymptotic infinity, with $r=0$ and $t=0$ being the bifurcation point of the Killing horizon. Increasing the acceleration magnitude $|a|$ in the latter case, that is for the Rindler trajectory in flat spacetime, one can bring the turning point $r_b^{rindler} = 1/|a|$ to be arbitrary closer to $r = 0$ or to the Rindler horizon at $t-r=0$ by increasing $|a|$ all the way upto infinity. In the present case, one has introduced a black hole centred at $r=0$. Here too, one would have expected in general, the $r_{min}$ to go all the way to the Schwarzschild radius $r_s$ for a continuous increase in the value of acceleration. The lower bound on $r_{min}$ is still inversely proportional to $|a|$ through $r_b = 1/(\sqrt{3} |a|)$ in Eq.(\ref{3.c-extremum at bound}). However, as demonstrated, increasing the acceleration $|a|$ beyond the bound $ |a| \leq 1/(\sqrt{27} M)$, simply thrusts the trajectory into the black hole horizon on crossing the lower bound radius $r_{min} = 3M$. One caveat to be noted is that near the saturation value, when $|a|$ is of the order of $1/(\sqrt{27} M)$, the  presence of the LUA trajectory may back-react on the background curvature which may lead to additional effects on the bounds. Further we have shown that a finite bound on the value of acceleration, $ |a| \leq B(M,h) = a_2$ given in Eq.(\ref{3.accsol}) and a corresponding distance of closest approach $r_{b} > 2M$ given in Eq.(\ref{3.bound}) always exists, for all finite asymptotic initial data $h$.
 
A similar bound $ \alpha \leq 1/(\sqrt{27} M)$ is known in the literature for a case of an uniformly accelerating black hole with magnitude of acceleration $\alpha$ as described by the C-metric \cite{cmetric}
\begin{eqnarray}
ds^{2}&=&\frac{1}{{(1+\alpha r\cos\theta)}^{2}}\left(-Q dt^{2}+\frac{dr^{2}}{Q}+\frac{r^{2}d{\theta}^{2}}{P}+Pr^{2}\sin^{2}\theta d{\phi}^{2}\right)
\label{cmetric}
\end{eqnarray}
where $P=(1+2\alpha m\cos\theta)$ and $Q=(1-{\alpha}^{2}r^{2})\left(1-2m/r\right)$. The C-metric spacetime has two horizons, the black hole horizon and the Rindler horizon corresponding to the two real distinct positive roots of $Q$ provided the bound on acceleration $\alpha$ is satisfied. At the  value $ \alpha = 1/(\sqrt{27} M)$ the bound is saturated and the two roots coincide, that is, the two horizons coincide. For any $\alpha$ greater than the saturation value, there is no horizon. In the present case of the LUA observer in the Schwarzschild metric, the roots of $u^1$ show the same algebraic behaviour, as explained in section \ref{sec3}, with the two real positive roots being the $r_{min}$ and $r_{max}$ of the LUA trajectory in Eq.(\ref{3.c-extremum1}) and Eq.(\ref{3.c-extremum3}). The similarity is because the polynomial under the square-root in $u^1$ in Eq.(\ref{3.velocity1}) is identical to the polynomial $Q$ in the C-metric Eq.(\ref{cmetric}). 

Hence, although the two cases (i) a uniformly accelerating black hole and (ii) a linearly uniformly accelerating observer in a static Schwarzschild black hole spacetime, describe two different physical scenarios, the equivalence between the bound relations brings up the question whether something curious happens with the Rindler horizon of the LUA observer in the latter case as well. As motivated in section 1, a full general relativistic treatment is needed to understand how the 2-D null hyper-surface of the Rindler horizon behaves close to the black-hole. We investigate this feature in a future publication \cite{kajol}.

In an earlier work \cite{kolekar2, kolekar3}, the quantum field aspects for a uniformly accelerated observer moving in an inertial thermal bath were investigated. It was shown that the reduced density matrix for the Rindler observer in a flat spacetime moving in an inertial thermal bath with temperature $T_{b}$ (instead of the usual inertial vacuum) with acceleration $a = 2 \pi T_{u}$, where $T_{u}$ is the Rindler horizon temperature, is symmetric in $T_{u}$ and $T_{b}$. Hence, it was argued that the Rindler observer is unable to distinguish between thermal and quantum fluctuations. The radially moving LUA observer in Schwarzschild spacetime is analogous to a Rindler observer moving in an existing thermal bath, which in this case is the Hawking thermal bath of the black hole. Thus an investigation into the quantum treatment of fields in the background Schwarzschild spacetime from the perspective of the LUA observer would be interesting \cite{kajol}.

\section*{Acknowledgments}

We thank T. Padmanabhan, Jorma Louko, Andreas Finke for a helpful 
discussion on uniformly accelerated trajectories in the Schwarzschild spacetime and for useful comments on the draft. SK and KP thank the Department of Science and Technology, India, for financial support. 

\section*{Appendix A} \label{Appen}

The solutions in Eqs.(\ref{2.result1}) and (\ref{2.result0}) can also be written by choosing the constant $c = |a| h$ as,
\begin{eqnarray}
u^{1}&=&\pm\sqrt{\left(|a|r+\frac{c}{|a|}\right)^2-f(r)}\label{2.c1}\\
u^{0}&=&{f(r)}^{-1}\left(|a|r+\frac{c}{|a|}\right)\label{2.c0}
\end{eqnarray}

For the general case, $c\neq 0$, we have
\begin{eqnarray}
\dfrac{dr}{dt}&=&\frac{u^{1}}{u^{0}}=\pm(1-\frac{r_{s}}{r})\frac{\sqrt{\left(|a|r+\frac{c}{|a|}\right)^2-1+\frac{r_{s}}{r}}}{\left(|a|r+\frac{c}{|a|}\right)}
\end{eqnarray}
Solving this differential equation numerically, the trajectory $r$ as a function of $t$ is plotted to get a curve  similar to that of figure(\ref{fig2}).
The extrema of this curve r(t) are obtained by solving $dr/dt=0$, to get the following three roots,

\begin{eqnarray}
r_{1}&=&-\frac{2c}{3{|a|}^2}+\frac{P^{1/3}}{2^{1/3}\:3{|a|}^{4}}+\frac{2^{1/3}(3{|a|}^2+c^2)}{3\:P^{1/3}}\\
r_{2}&=&-\frac{2c}{3{|a|}^{2}}-\frac{(1-i \sqrt{3})P^{1/3}}{{2}^{1/3}\:6{|a|}^{4}}-\frac{(1+i\sqrt{3})(3{|a|}^{2}+c^{2})}{2^{2/3}\:3\:P^{1/3}}\\
r_{3}&=&-\frac{2c}{3{|a|}^{2}}-\frac{(1+i \sqrt{3})P^{1/3}}{{2}^{1/3}\:6{|a|}^{4}}-\frac{(1-i\sqrt{3})(3{|a|}^{2}+c^{2})}{2^{2/3}\:3\:P^{1/3}}
\end{eqnarray}

here,
$P=-A+iB$, with, $A=(27{|a|}^{10}r_{s}+18{|a|}^{8}c-2{|a|}^{6}c^{3})$ and $B=\sqrt{4(3{|a|}^{6}+{|a|}^{4}c^{2})^{3}-(A)^{2}}$. Simplifying these expressions in a similar manner as in the case $h=0$, we get,

\begin{eqnarray}
r_{1}&=&\frac{2}{3{|a|}^{2}}\Big[\frac{\sqrt{3{|a|}^{2}+c^{2}}}{2}\left(\cos(\xi/3)+\sqrt{3}\sin(\xi/3)\right)-c\Big]\label{3.extremum1}\\
r_{2}&=&\frac{-2}{3{|a|}^{2}}\Big[\sqrt{3{|a|}^{2}+c^{2}}\cos(\xi/3)+c\Big]\label{3.extremum2}\\
r_{3}&=&\frac{2}{3{|a|}^{2}}\Big[\frac{\sqrt{3{|a|}^{2}+c^{2}}}{2}\left(\cos(\xi/3)-\sqrt{3}\sin(\xi/3)\right)-c\Big]\label{3.extremum3}
\end{eqnarray}

where, $\xi=\arctan(B/A)$. For $r_{1}$, $r_{2}$ and $r_{3}$ to be real, $B$ needs to be real, that is, $4(3{|a|}^{6}+{|a|}^{4}c^{2})^{3}-(A)^{2}$ should be greater than or equal to zero, and thus $\xi$ will be in the range $0<\xi<\pi$. It is then clear from the equations (\ref{3.extremum1}), (\ref{3.extremum2}), (\ref{3.extremum3}) that, $r_{2}$ will always be negative since $\cos(\xi/3)$ is positive, and thus it does not represent a physical solution. The root $r_{1}$ gives us the minimum of the trajectory.

To find the range of acceleration for which, $B$ is real, i.e $B^{2}\geq 0$, we first obtain the values of acceleration $|a|$ which give the zeros of function $B$, that is imposing $B=0$, the acceleration $|a|$ is obtained in terms of $c$ and $r_{s}$. The non-negative real solutions are,
\begin{eqnarray}
a_{0}&=&0 \nonumber \\
a_{1}&=&\sqrt{\frac{4}{81{r_{s}}^{2}}-\frac{4c}{9r_{s}}+\frac{2^{2/3}N}{Q^{1/3}}+\frac{2^{1/3}Q^{1/3}}{81{r_{s}}^{2}}} \nonumber \\
a_{2}&=&\sqrt{\frac{4}{81{r_{s}}^{2}}-\frac{4c}{9r_{s}}+\frac{2^{2/3}N}{2Q^{1/3}}(1-\sqrt{3}i)-\frac{Q^{1/3}}{2^{2/3}81{r_{s}}^{2}}(1+\sqrt{3}i)} \nonumber\\
a_{3}&=&\sqrt{\frac{4}{81{r_{s}}^{2}}-\frac{4c}{9r_{s}}+\frac{2^{2/3}N}{2Q^{1/3}}(1+\sqrt{3}i)-\frac{Q^{1/3}}{2^{2/3}81{r_{s}}^{2}}(1-\sqrt{3}i)}
\label{accsol}
\end{eqnarray}
where, $N=8/81{r_{s}}^{2}-16c/9r_{s}+4c^{2}+2c^{3}r_{s}$ and $Q=C+D$, with, $C=10935\,c^{4}{r_{s}}^{4}-4860\,c^{3}{r_{s}}^{3}+5832\,c^{2}{r_{s}}^2-864\,cr_{s}+32$
 and  $D=81\sqrt{3}\sqrt{M}$ 
with $M=-432\,c^{9}{r_{s}}^{9}+3483\,c^{8}{r_{s}}^{8}-9432\,c^{7}{r_{s}}^{7}+8768\,c^{6}{r_{s}}^{6}-512\,c^{5}{r_{s}}^{5}$

It can be checked that, when $D$ is real ($M\geq0$) then $Q$ becomes real, and it is possible to have a real non-negative solution $a_{1}$ with $a_{2}$, $a_{3}$ to be complex for a certain range of $c$. While for an imaginary $D$, it is possible to have all the solutions $a_{1}$, $a_{2}$ and $a_{3}$ to be non-negative real or just $a_{1}$ to be non-negative real  and $a_{2}$, $a_{3}$ to be complex for a certain range of $c$. Below, we find those values of $c$. 

Using the same method as done for the radii $r_{1}$, $r_{2}$ and $r_{3}$, the solutions in Eq.(\ref{accsol}) can be re-expressed as
\begin{eqnarray}
a_{1}&=&\sqrt{\frac{4}{81{r_{s}}^{2}}-\frac{4c}{9r_{s}}+\frac{2{(2R)}^{1/3}}{81{r_{s}}^{2}}\cos(\frac{\beta}{3})}\label{eq81}\\
a_{2}&=&\sqrt{\frac{4}{81{r_{s}}^{2}}-\frac{4c}{9r_{s}}-\frac{2{(2R)}^{1/3}}{81{r_{s}}^{2}}\cos(\frac{\pi+\beta}{3})}\label{eq82}\\
a_{3}&=&\sqrt{\frac{4}{81{r_{s}}^{2}}-\frac{4c}{9r_{s}}-\frac{2{(2R)}^{1/3}}{81{r_{s}}^{2}}\cos(\frac{\pi-\beta}{3})}\label{eq83}
\end{eqnarray}
where, $\beta=\arctan(D^{'}/C)$, $D^{'}=81\sqrt{3}\sqrt{M^{'}}$ and $M^{'}=-M$.
Now, to have all the real positive solutions, we need to have $D^{'}$ real, that is, $M^{'}$ should be positive or zero.
To find the range of $c$ for which $M^{'}$ is greater than or equal to zero, we first find the zeros of $M^{'}$. Solving $M^{'}=0$, for $c$, we get, $c=0$, or $c=1/16r_{s}$ or $c=8/3r_{s}$.

\begin{figure}[h]
\centering
\begin{minipage}{.5\textwidth}
\includegraphics[width=7.5cm,height=5.5cm]{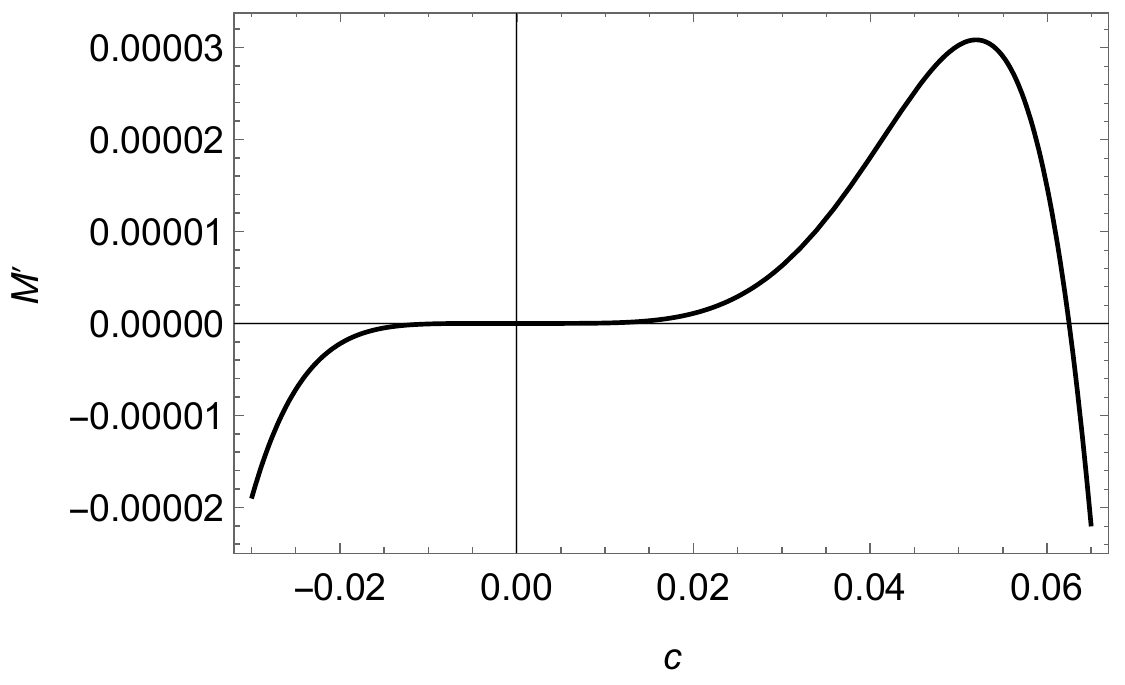}
\end{minipage}%
\begin{minipage}{.5\textwidth}
\includegraphics[width=7.5cm,height=5.5cm]{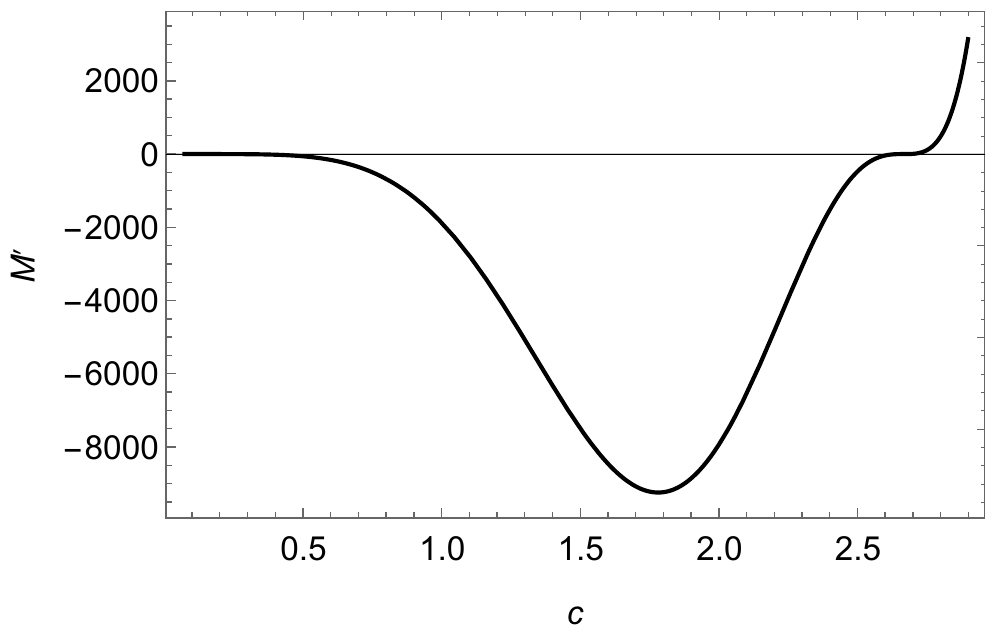}
\end{minipage}
\caption{$M^{'}$ as a function of $c$ for $r_{s}=1$. The function has zeros at, $c=0$, $c=0.0625$ and $c=2.6667$.}
\label{fig3}
\end{figure}

One can check that $M^{'}\geq0$ for the range $0\leq c\leq (1/16r_{s})$ and $c\geq (8/3r_{s})$ while $M^{'}<0$ for $c<0$ and the range $(1/16r_{s})<c<(8/3r_{s})$. A Plot of $M^{'}$ as a function of $c$ for a particular value of $r_{s}=1$ is shown in figure(\ref{fig3}).

The values for $a_{1}$, $a_{2}$ and $a_{3}$ are calculated for different values of $c$ in each range of $c$ and it was found that, in range $0 \leq c \leq (1/16r_{s})$, $a_{1}$, $a_{2}$ and $a_{3}$ are all real and positive, but for $c\geq (8/3r_{s})$, only $a_{1}$ is real and positive and $a_{2}$ and $a_{3}$ are complex values and thus are not physical solutions.
For the other two ranges of $c$, i.e, for $c<0$ and $(1/16r_{s})<c<(8/3r_{s})$, only $a_{1}$ is real and positive, and $a_{2}$ and $a_{3}$ are complex values.

\begin{figure}[h]
\centering
\begin{minipage}{.5\textwidth}
\includegraphics[width=7.5cm,height=5.5cm]{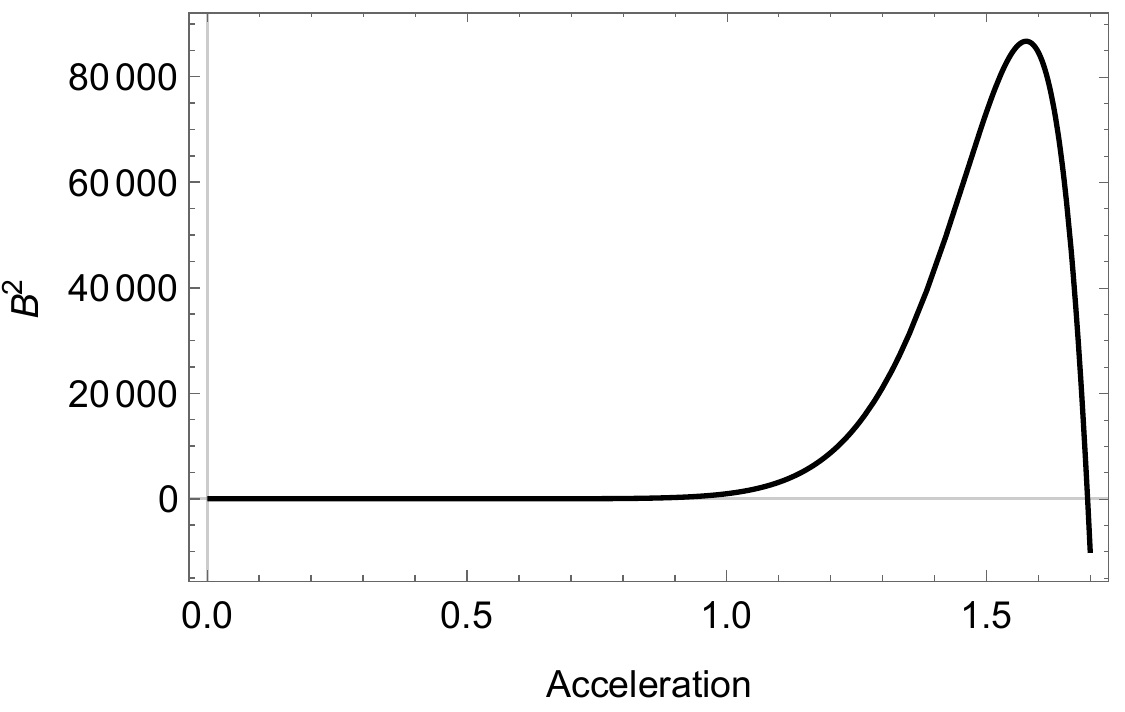}
\end{minipage}%
\begin{minipage}{.5\textwidth}
\includegraphics[width=7.5cm,height=5.5cm]{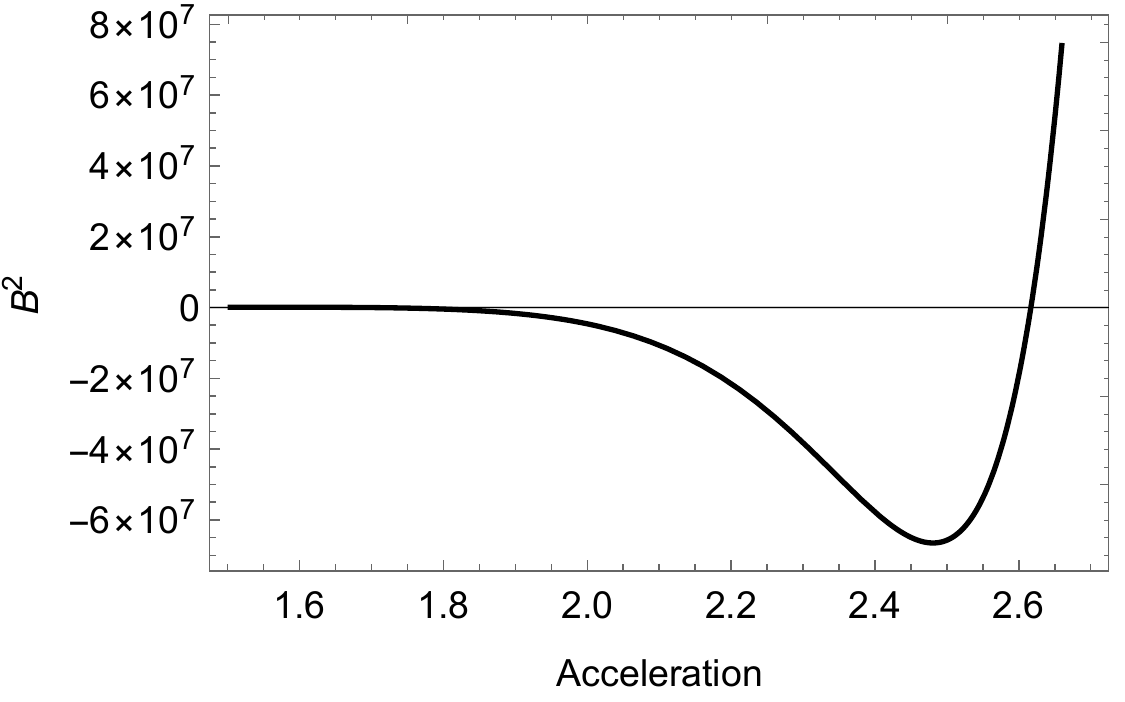}
\end{minipage}
\includegraphics[width=7.5cm,height=5.5cm]{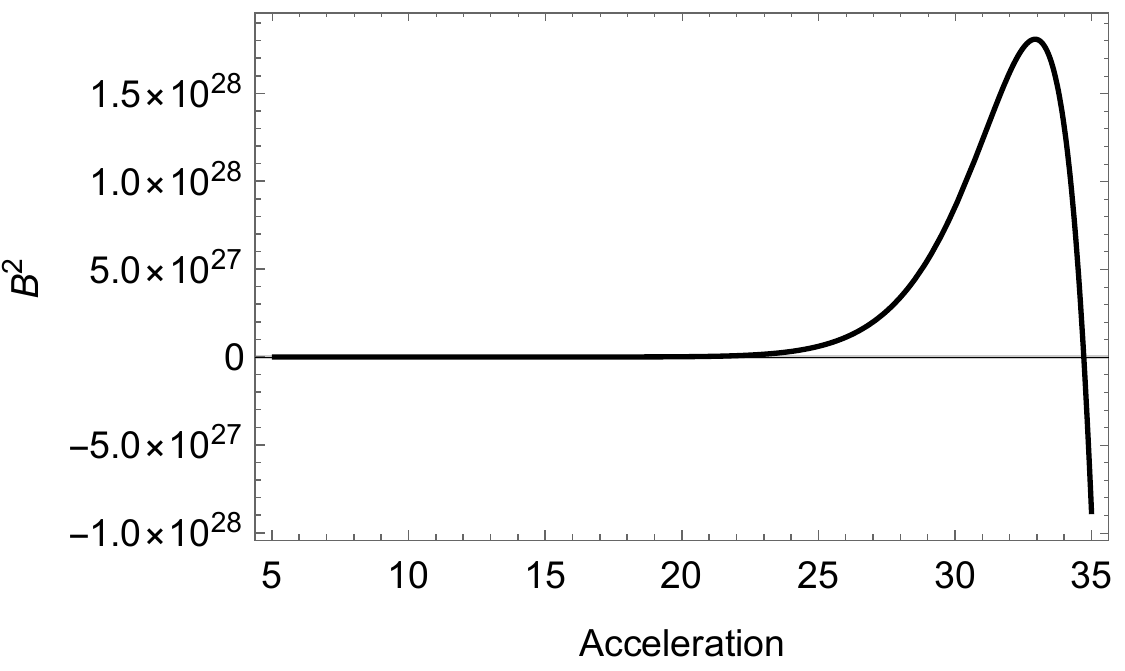}
\caption{Plot of $B^2 = 4(3 {|a|}^6+{|a|}^4 c^2)^{3}-(27{|a|}^{10}r_{s}+18{|a|}^{8}c-2{|a|}^{6}c^{3})^{2}$ as a function of $|a|$ for $c=2$ and $r_{s}=0.01$. The zeros of the function are $a_{0}=0$, $a_{1}=34.7145$, $a_{2}=2.61681$ and $a_{3}=1.69483$.}
\label{fig4}
\end{figure}

\begin{figure}[h!]
\begin{subfigure}{.5\textwidth}
\includegraphics[width=7.5cm,height=5.5cm]{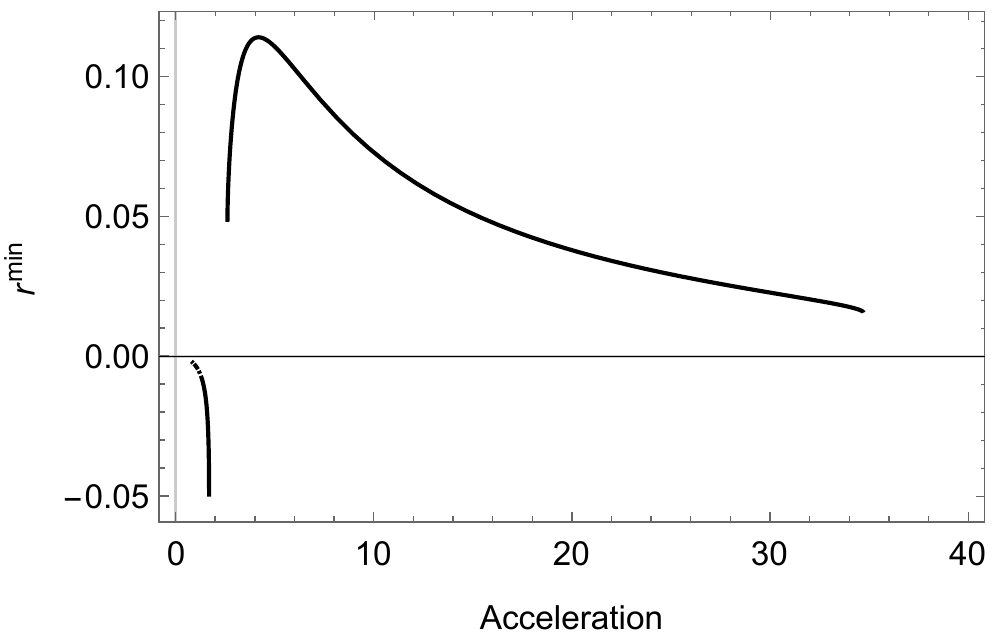}
\caption{$r_{min}$}
\end{subfigure}
\begin{subfigure}{.5\textwidth}
\includegraphics[width=7.5cm,height=5.5cm]{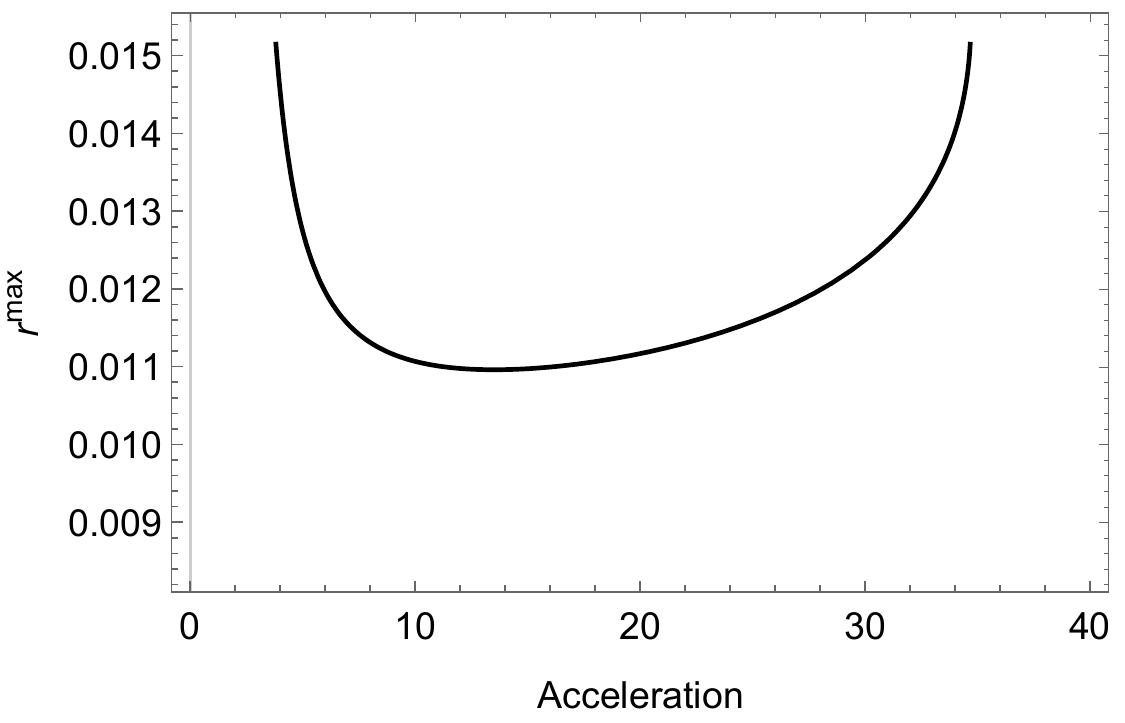}
\caption{$r_{max}$}
\end{subfigure}
\caption{$r_{1} = r_{min}$ and $r_{3} = r_{max}$ as a function of $|a|$ for $c=2$ and $r_{s}=0.01$}
\label{fig5}
\end{figure}

Now, as stated earlier, to have the minimum $r_{1}$ and the maximum $r_{3}$ real, the term $4(3 {|a|}^6+{|a|}^4 c^2)^{3}-(27{|a|}^{10}r_{s}+18{|a|}^{8}c-2{|a|}^{6}c^{3})^{2}$ should be greater than or equal to zero. To check the range of magnitude of acceleration $|a|$ for which this term is positive, it is plotted as a function of $|a|$ (the plot for $c=2$ and $r_{s}=0.01$ is shown in figure \ref{fig4}) and the extrema, $r_{1}$ and $r_{3}$ are also plotted as a function of $|a|$ (see figure \ref{fig5}) for a particular value of $c$ in each range of $c$.

The nature of solutions $a_{1}$, $a_{2}$ and $a_{3}$ and $r_{1}$ and $r_{3}$ in each range of $c$ is tabulated below in table(\ref{Table_1}).

\begin{table}[h!]
\centering
\begin{tabular}{|c|c|c|c|}
\hline
Range of $c$ & Nature of $a_{1}$, $a_{2}$, $a_{3}$ & Range of acceleration $a$ & Nature of $r_{1}$ and $r_{3}$\\
\hline
$c<0$ & $a_{1}\to real,$ & $0<a\leq a_{1}$ & $real$ and $positive$\\
\cline{3-4}
 & $a_{2}$, $a_{3}\to complex$ & $a>a_{1}$ & $complex$ \\
\hline
$0\leq c\leq \frac{1}{16r_{s}}$ & $a_{1}$,$a_{2}$,$a_{3}\to real$ & $0<a<a_{2}$ & $complex$ or $negative$ \\
\cline{3-4}
 & $a_{3}<a_{2}<a_{1}$ & $a_{2}\leq a \leq a_{1}$ & $real$ and $positive$\\
\cline{3-4}
 & & $a>a_{1}$ & $complex$ \\
\hline
$\frac{1}{16r_{s}}<c<\frac{8}{3r_{s}}$ & $a_{1}\to real,$ & $0<a\leq a_{1}$ & $complex$ or $negative$ \\
\cline{3-4}
 & $a_{2}$, $a_{3}\to complex$ & $a>a_{1}$ & $complex$ \\
\hline
$c\geq \frac{8}{3r_{s}}$ & $a_{1}\to real,$ & $0<a\leq a_{1}$ & $complex$ or $negative$ \\
\cline{3-4}
 & $a_{2}$, $a_{3}\to complex$ & $a>a_{1}$ & $complex$ \\
\hline
\end{tabular}
\caption{Nature of solutions $a_{1}$, $a_{2}$ and $a_{3}$ and $r_{1}$ and $r_{3}$.}
\label{Table_1}
\end{table}

From the table, one can see that, $r_{1}$ and $r_{3}$ are positive and real only for the range $|a|\leq a_{1}$ of acceleration in the range $c<0$ and for the range $a_{2}\leq |a|\leq a_{1}$ of acceleration in the range $0\leq c\leq (1/16r_{s})$.
Thus to have a radial  LUA trajectory with a turning point in Schwarzschild metric it should have acceleration in range $|a|\leq a_{1}$ when $c<0$ and in range $a_{2}\leq |a|\leq a_{1}$ when $0\leq c\leq (1/16r_{s})$. In all other cases, the LUA trajectory initially moving towards the black hole falls into the black hole horizon.

\end{document}